\documentclass[12pt]{article}
\usepackage{amsmath}
\usepackage{graphicx,psfrag,epsf}
\usepackage{enumerate}
\usepackage{natbib}

\usepackage{color}

\usepackage[toc,page]{appendix} 
\usepackage{dsfont}

\newcommand{\blind}{0}

\begin{document}

\def\spacingset#1{\renewcommand{\baselinestretch}%
{#1}\small\normalsize} \spacingset{1}


\if0\blind
{
  \title{\bf Model-based clustering for conditionally correlated categorical data}
  \author{Matthieu Marbac\thanks{Matthieu Marbac is a Ph.D student at Inria Lille  and DGA (Email: matthieu.marbac-lourdelle@inria.fr)}\hspace{.2cm}\\
   Inria \&  DGA\\
\and Christophe Biernacki\thanks{Christophe Biernacki is a Professor at University Lille 1, CNRS and Inria Lille (Email: Christophe.Biernacki@math.univ-lille1.fr)} \hspace{.2cm}\\
 University Lille 1 \& CNRS \& Inria \\
\and Vincent Vandewalle\thanks{Vincent Vandewalle is an Associate Professor at University Lille 2, EA 2694 and Inria Lille (Email: vincent.vandewalle@univ-lille2.fr)}\hspace{.2cm}\\ EA 2694  University Lille 2  \& Inria \\ 
}
  \maketitle
} \fi

\if1\blind
{
  \bigskip
  \bigskip
  \bigskip
  \begin{center}
    {\LARGE\bf  Model-based clustering for conditionally correlated categorical data }
\end{center}
  \medskip
} \fi

\bigskip
\begin{abstract}
An extension of the latent class model is presented for clustering categorical data by relaxing the classical ``class conditional independence assumption'' of variables. This model consists in grouping the variables into inter-independent and intra-dependent blocks, in order to consider the main intra-class correlations. The dependency between variables grouped inside the same block of a class is taken into account by mixing two extreme distributions, which are respectively the independence and the maximum dependency. When the variables are dependent given the class, this approach is expected to reduce the biases of the latent class model. Indeed, it produces a meaningful dependency model with only a few additional parameters. The parameters are estimated, by maximum likelihood, by means of an EM algorithm. Moreover, a Gibbs sampler is used for model selection in order to overcome the computational intractability of the combinatorial problems involved by the block structure search. Two applications on medical and biological data sets show the relevance of this new model. The results strengthen the view that this model is meaningful and that it reduces the biases induced by the conditional independence assumption of the latent class model.
\end{abstract}

\noindent%

\spacingset{1.45}

\section{Introduction}

Nowadays practitioners often face very large data sets, which are difficult to analyze directly. In this context, clustering  \citep{Jaj02} is an important tool providing a partition among individuals. Other approaches cluster both individuals and variables simultaneously \citep{Gov03}. Furthermore, with the increasing number of variables at hand, the risk of observing correlated descriptors, even within the same class, is often high. In view of these difficulties, the practitioner can choose between two approaches. The first approach consists in performing a selection among the observed variables \citep{Mau09} in order to extract uncorrelated data, thereby losing some potentially crucial information. The second approach consists in applying a method for modeling the conditional dependencies on the whole set of variables. \bigskip

Clustering methods can be divided into two kinds of approaches: geometric ones based on the distances between individuals and probabilistic ones modelling the data generation process. Geometric methods are generally faster than probabilistic ones, but they are often quite sensitive to the choice of distance between individuals. Furthermore, as the probabilistic tools are not available for these methods, difficult questions, e.g. the selection of the number of clusters cannot be addressed rigorously. For categorical data, geometric methods either define a metric in the initial variables space e.g. \emph{k-means} \citep{Hua05}, or compute their metric on the axes of the multiple correspondence analysis \citep{Cha10,Gui01}.

Many geometric approaches can also be interpreted in a probabilistic way. Thus, for the continuous data, the classical \emph{k-means} algorithm can be identified as an homoscedastic Gaussian mixture model \citep{Ban93,Cel95} with equal proportions. For the categorical variables, \citet{Cel91} show that the CEM algorithm \citep{Mcl97}, applied to a classical latent class model, maximizes a classical information criteria close to a $\chi^2$ metric. Other links between the two approaches are described in \citet{Gov10}, Chapter 9. Let us now introduce our proposal for this problem.
\bigskip

In the categorical case, the \emph{latent class model} also known as naive Bayes belongs to the folklore \citep{Goo74,Cel91}. In this article, we refer to this model as the conditional independence model (further denoted by \textsc{cim}). Classes are explicitly described by the probability of each modality for each variable under the conditional independence assumption. The sparsity of the model, implied by this assumption, is a great advantage since it restricts the curse of dimensionality. \textsc{cim} has obtained quite good results in practice in different areas \citep{Han01}, e.g. behavioral science \citep{Reb06} and medicine \citep{Str06}. However, \textsc{cim} may suffer from severe biases when the data are intra-class correlated. For instance, an application presented by \citet{Van09} shows that \textsc{cim} over-estimates the number of clusters when the conditional independence assumption is violated. For a long time, people have tried to relax the conditional independence assumption by modeling conditional interactions between variables using an additive model \citep{Har72}. The main drawback of this approach is that its number of parameters becomes huge, making their estimation intractable.

Some other methods take into account the intra-class dependencies as \emph{mixtures of Bayesian networks} \citep{Che99}. Conditionally on each class, a directed acyclic graph is built with a set of nodes representing each variable. However, if no constraint is added, the network estimation is also quite complex. By constraining the network to be a tree, the model selection and the parameter estimation can be easily performed. Moreover the correlation model enjoys great flexibility. The extension of the dependency tree of \citet{Cho68} was done by \citet{Fri97} for the supervised classification and by \citet{Mei01} for the clustering. However the main problem of these models is that they too often require too many parameters.
 
When covariates are available, the conditional dependencies between the categorical ones can be modeled by a logistic function \citep{For92,Reb08}. By assuming that these covariates are unobserved, the \emph{multilevel latent class model} \citep{Ver03,Ver07} naturally incorporates the intra-class dependencies. This model has connections with the approach of \citet{Qu96} where the intra-class dependencies are modeled by a latent continuous variable with a probit function. The \emph{hybrid model} \citep{Mut08} is  more general approach in which, for each class, a factor analysis model is fitted to either all categorical variables or to those categorical variables having dependencies. Recently, \citet{Gol13} proposed the \emph{mixture model of latent traits analyzers} which assumes that the distribution of the categorical variables depends on both a categorical latent variable (the class) and many continuous latent traits variables. The parameter estimation is also a difficult point which is solved via a variational approach. All these models consider the intra-class dependencies, but their main drawback is that these dependencies must be interpreted among relations with a latent variable. As a result, pertinent interpretation can be difficult.

The log-linear models \citep{Agr02,Boc86} were originally proposed to model the log-probability of each individual by selecting interactions between variables. The most general mixture model is the \emph{log-linear mixture model} as it is able to incorporate many forms of interactions. It has been used at least since \citet{Hag88}. \citet{Esp89} used it to cluster radiographic cross-diagnostics and \citet{Van09} in a market segmentation problem. However this model family is huge and the model selection is a real challenge. In the literature, authors often require the modeled interactions  ahead of time. Another option is to perform a deterministic search e.g. the \emph{forward} method which is sub-optimal. Furthermore, the number of parameters to estimate increases with the conditional modalities interactions, thus implying potential over-fitting and more difficult interpretation. The latent class model (\textsc{cim}) can be seen as a particular log-linear mixture model, in which interactions are discarded. Our aim is to present a version of the log-linear mixture model which takes into account the interactions of order one or more while keeping the number of unknown parameters to a reasonable amount.
\bigskip

We propose to extend the classical latent class model (\textsc{cim}) for categorical data, by a new latent class model which relaxes the conditional independence assumption. We refer to this new model as the \emph{conditionally correlated model} (denoted by \textsc{cmm}). This model is a parsimonious version of the log-linear mixture model, and thus benefits from its interpretative power. Furthermore, we propose a Bayesian approach to automatically perform model selection.

The \textsc{ccm} model groups the variables into conditionally independent blocks given the class. The main intra-class dependencies are thus shown by the repartition of the variables into blocks. This approach, allowing modeling of the main conditional interactions, was first proposed by \citet{Jor96} in order to cluster continuous and categorical data. For \textsc{cmm}, each block follows a particular dependency distribution which corresponds to our main contribution. This distribution consists in a bi-component mixture of an \emph{independence} and a \emph{maximal dependency} distribution according to the Cramer's V criterion. This specific distribution of the blocks allows summarizing the conditional dependencies of the variables  with only one continuous parameter: the maximum dependency distribution proportion. Thus, the model underlines the main conditional dependencies and their strength.

The new model is a two-degree parsimonious version of a log-linear mixture model. A first degree of parsimony is introduced by grouping the variables which are conditionally dependent into the same block. This repartition of the variables per blocks defines the interactions considered by the model for each class. Moreover, the strength of the correlation is reflected by the proportion of maximum dependency distribution. A second degree of parsimony is induced by the specific distribution of the blocks. As for all log-linear mixture models, the selection of the pertinent interactions is a combinatorial problem. We propose to perform this model selection via a Gibbs sampler in order to overcome the enumeration of all the models. Thus, this general approach could also select the interactions of any log-linear mixture model.

\bigskip

This paper is organized as follows. Section~\ref{latentclass} reviews the principles of the latent class model. Section~\ref{ourmodel} presents the new mixture model taking into account the intra-class correlations. Section~\ref{estimation} is devoted to parameter estimation in the case where the number of classes and the blocks of variables are supposed to be known. Section~\ref{modelsel} presents a Gibbs algorithm for avoiding the combinatorial difficulties inherent to block selection. Section~\ref{simulations} presents results on simulated data. Section~\ref{applications} presents a comparison between two main model-based clustering approaches and our proposal on a classical medical data set, and presents another application on a larger real data set. A tutorial of the R package \texttt{Clustericat}\footnote{The R package \texttt{Clustericat} is available on Rforge website at the following url: \emph{https://r-forge.r-project.org/R/?group\_id=1803}} which performs the model selection and the estimation of the parameters of \textsc{cmm} is given with the first application (see Appendix~\ref{tuto}). Section~\ref{conclu} presents our conclusions.

\section{Classical models\label{latentclass}}

\subsection{Latent class model: intra-class independence of variables}
Observations to be classified are described with $d$ discrete variables $\boldsymbol{x}=(\boldsymbol{x}^1,\ldots,\boldsymbol{x}^d)$ defined on the probabilistic space $\mathcal{X}$. Each variable $j$ has $m_{j}$ response levels with $m_j\geq 2$ and is written $\boldsymbol{x}^j=(x^{j1},\ldots,x^{jm_j})$ where $x^{jh}=1$ if variable $j$ takes modality $h$ and $x^{jh}=0$ otherwise.  In the standard latent class model (\textsc{cim}), the variables are assumed to be \emph{conditionally independent} knowing the latent cluster. Furthermore, data are supposed to be drawn independently from a mixture of $g$ multivariate multinomial distributions with probability distribution function (pdf)
	
	\begin{equation}
		p(\textbf{\textit{x}} ; \boldsymbol{\theta})=\sum_{k=1}^{g} \pi_{k}  \mathring{p}(\textbf{\textit{x}} ;\boldsymbol{\alpha}_{k})\qquad \text{ with } \qquad \mathring{p}(\textbf{\textit{x}};\boldsymbol{\alpha}_{k})=\prod_{j=1}^{d}\prod_{h=1}^{m_{j}}(\alpha_{k}^{jh})^{x^{jh}},
	\end{equation}
with $\boldsymbol{\theta}=(\pi_1,\ldots,\pi_g,\boldsymbol{\alpha}_1,\ldots,\boldsymbol{\alpha}_g)$, $ \pi_k$ being the proportion of the component $k$ in the mixture where $\pi_k>0$ and $\sum_{k=1}^g \pi_k=1$, and $\boldsymbol{\alpha}_{k} =(\alpha_{k}^{jh};j = 1, \ldots , d;h = 1,\ldots, m_{j})$ where $\alpha_{k}^{jh}$ denotes the probability that variable $j$ has level $h $ if the object is in cluster $k$ and satisfies the two following constraints: $\alpha_{k}^{jh}>0$ and $\sum_{h=1}^{m_j}\alpha_k^{jh}=1$.

The classical latent class model  is much more parsimonious than the saturated log-linear model, which requires $(\prod_j m_j) -1$ parameters, since it only requires $\nu_{\textsc{cim}}$ parameters with
\begin{equation}
\nu_{\textsc{cim}}=(g-1)+g\sum_{k=1}^g (m_j-1). \label{nbparamcim}
\end{equation}
 Its maximum likelihood estimator is easily computed via an EM algorithm \citep{Mcl97}. For the cluster analysis, the mixture identifiability up to a permutation of the class is generally necessary \citep{Mcl00}. However, there are mixtures, such as the products of Bernoulli distributions, which are not identifiable but which produce good results in several  applications. In order to relax this too stringent concept of identifiability, the notion of generic identifiability was introduced by \citet*{All09}: a model is generically identifiable if it is identifiable except for a subset of the parameter space with Lebesgue measure zero.

\subsection{Latent class model extension: intra-class independence of blocks}\label{introex}
Despite its simplicity, the latent class model s to good results in many situations \citep{Han01}. However, in the case of intra-correlated variables, it can entail severe biases in the partition estimation and it may also overestimate the number of clusters. In order to reduce these biases, a classical extension of the latent class model  was introduced by \citet{Jor96} for conditionally correlated mixed data. This model is implemented in the Multimix software \citep{Hun99}.

It considers that \emph{conditionally} on the class $k$, variables are grouped into $\textsc{b}_k$ \emph{independent blocks} and that each block follows a specific distribution. The repartition in blocks of the variables determines a partition $\boldsymbol{\sigma}_k=(\boldsymbol{\sigma}_{k1},\ldots,\boldsymbol{\sigma}_{k\textsc{b}_k})$ of $\{1,\ldots,d\}$ in $\textsc{b}_k$ disjoint non-empty subsets where $\boldsymbol{\sigma}_{kb}$ represents the subset $b$ of variables in the partition $\boldsymbol{\sigma}_k$. This partition defines $\boldsymbol{x}^{\{kb\}}=\boldsymbol{x}^{\boldsymbol{\sigma}_{kb}}=(\boldsymbol{x}^{\{kb\}j};j=1,\ldots,d^{\{kb\}})$ which is the subset of $\boldsymbol{x}$ associated to  $\boldsymbol{\sigma}_{kb}$. The integer $d^{\{kb\}}=\text{card}(\boldsymbol{\sigma}_{kb})$ is the number of variables in block $b$ of component $k$ and $\boldsymbol{x}^{\{kb\}j}=({x}^{\{kb\}jh};h=1,\ldots,m_j^{\{kb\}})$ corresponds to variable $j$ of block $b$ for component $k$ with $x^{\{kb\}jh}=1$ if the individual takes modality $h$ for variable $\boldsymbol{x}^{\{kb\}j}$ and $x^{\{kb\}jh}=0$ otherwise and where $m_j^{\{kb\}}$ represents the number of modalities of $\boldsymbol{x}^{\{kb\}j}$. Note that different repartitions of the variables into blocks are allowed for each component and they are grouped into $\boldsymbol{\sigma}=(\boldsymbol{\sigma}_1,\ldots,\boldsymbol{\sigma}_g)$.

\begin{figure}[h!]
\begin{center}
\begin{minipage}{0.49\textwidth}
\centering \includegraphics[scale=0.4]{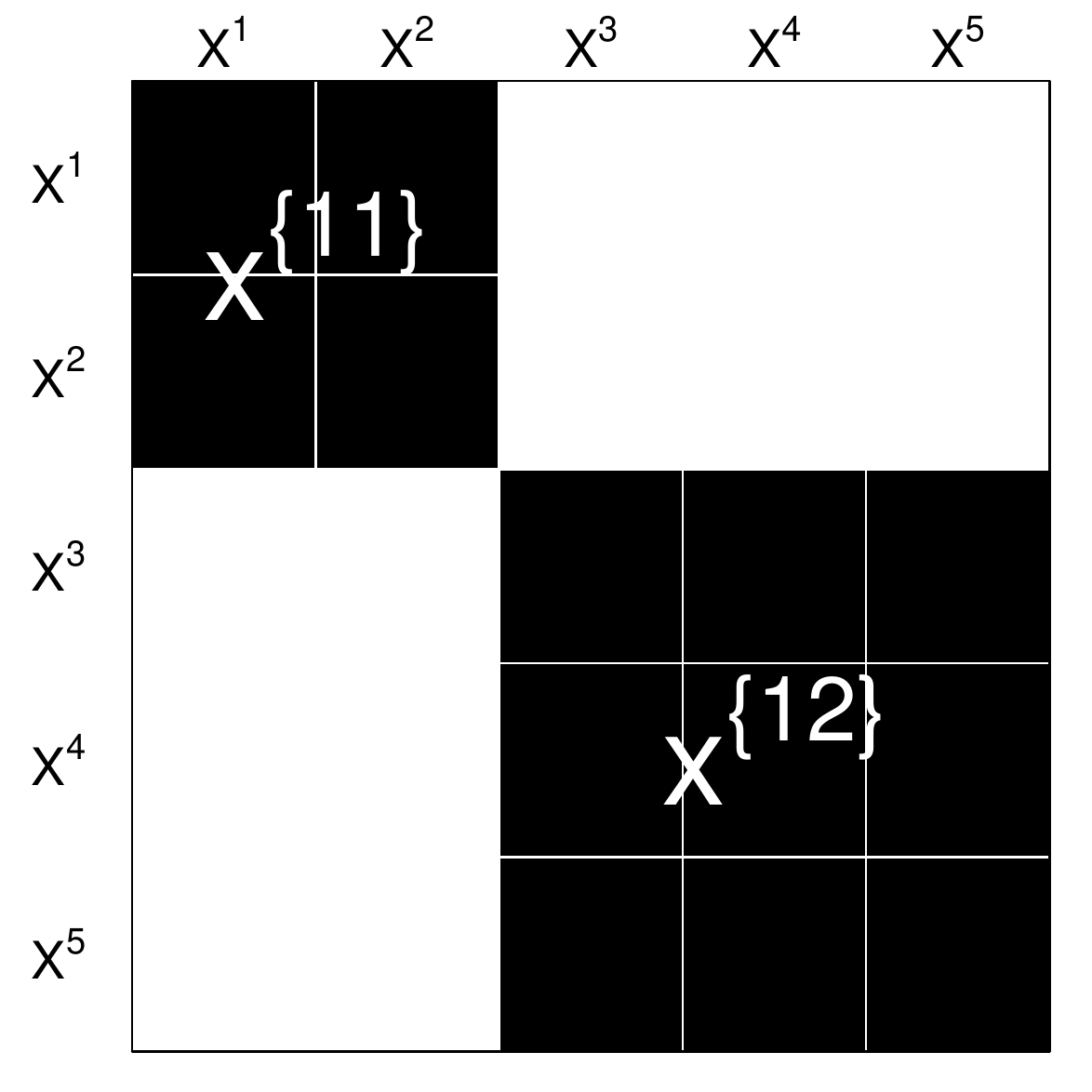} \\
\centering (a)
\end{minipage} 
 \hfill
\begin{minipage}{0.49\textwidth}
\centering \includegraphics[scale=0.4]{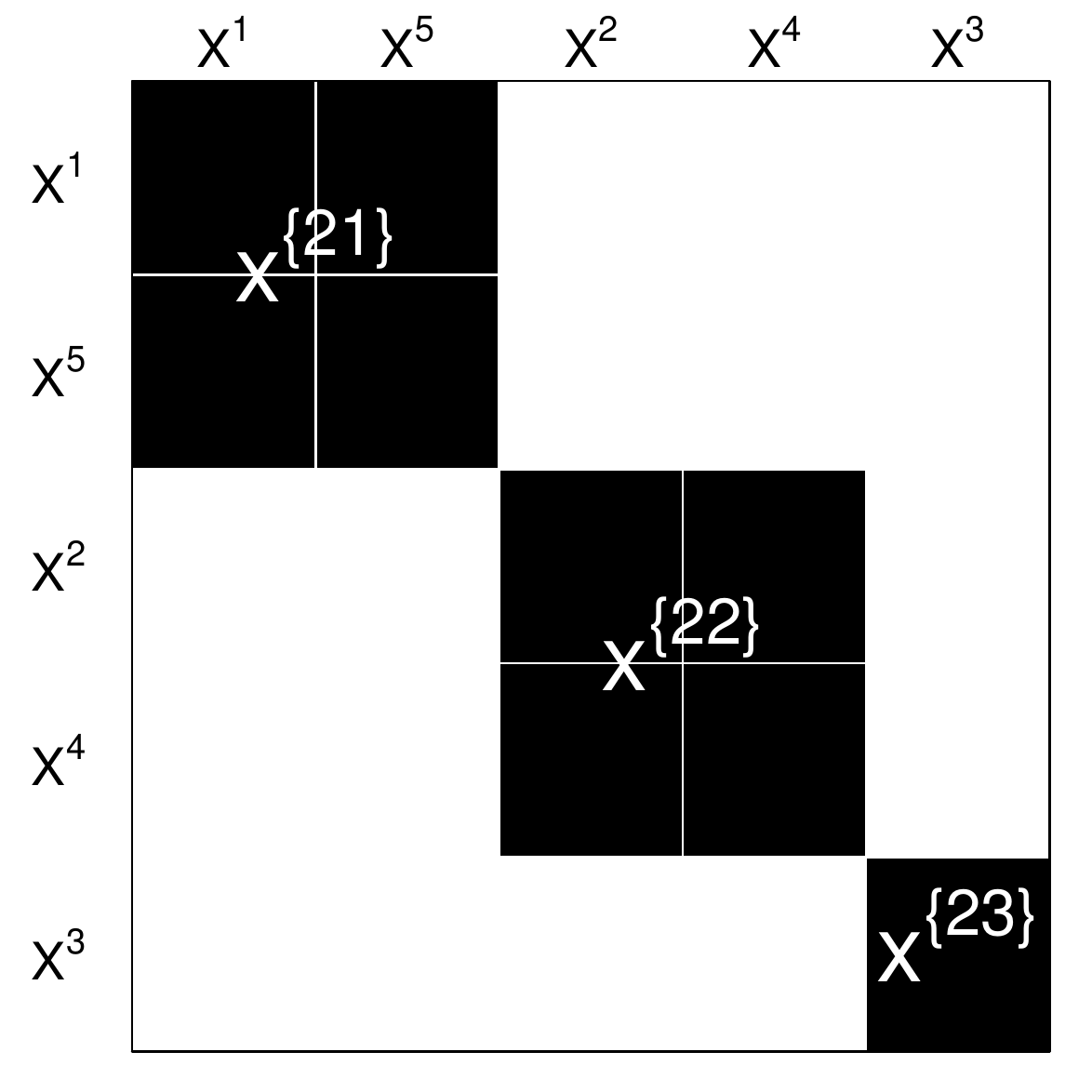} \\
\centering (b)
\end{minipage} 
 \end{center}
\centering \caption{Example of the  conditional independent blocks mixture model with $g=2$ and $d=5$, (a) $k=1$, $B_1=2$ and $\boldsymbol{\sigma}_{1}=(\{1,2\},\{3,4,5\})$, (b) $k=2$, $B_2=3$ and $\boldsymbol{\sigma}_{2}=(\{1,5\},\{2,4\},\{3\})$.}\label{exeblo}
 \end{figure}

For each component $k$, each block $b$ follows a specific parametric distribution denoted as $p(\textbf{\textit{x}}^{\{kb\}};\boldsymbol{\theta}_{kb}) $ where $\boldsymbol{\theta}_{kb}$ groups the parameters of this distribution. The model pdf is written as
\begin{equation}
p(\textbf{\textit{x}};\boldsymbol{\sigma},\boldsymbol{\theta})=
\sum_{k=1}^{g} \pi_{k} p(\textbf{\textit{x}};\boldsymbol{\sigma}_k,\boldsymbol{\theta}_{k})\qquad  \text{ with } \qquad p(\textbf{\textit{x}};\boldsymbol{\sigma}_{k},\boldsymbol{\theta}_{k})=\prod_{b=1}^{B_{k}}p(\textbf{\textit{x}}^{\{kb\}};\boldsymbol{\theta}_{kb}),
\end{equation}
 where $ \boldsymbol{\theta}$ is redefined as $ \boldsymbol{\theta}=(\pi_1,\ldots,\pi_g,\boldsymbol{\theta}_1,\ldots,\boldsymbol{\theta}_g)$ with $\boldsymbol{\theta}_k = (\boldsymbol{\theta}_{k1},\ldots,\boldsymbol{\theta}_{k\textsc{b}_k})$. Figure~\ref{exeblo} presents an example of the distribution with conditional independent blocks for a mixture with two components described by five variables. Blank cells indicate that the intra-class correlation is neglected and black cells indicate that this correlation is taken into account. For instance, Figure~\ref{exeblo}.a illustrates the distribution of the class 1 where two blocks ($\textsc{b}_1=2$) are considered. More precisely, the first block is composed by the first two variables ($\boldsymbol{\sigma}_{11}=\{1,2\}$) and the second block is composed by the last three variables ($\boldsymbol{\sigma}_{12}=\{3,4,5\}$). Note that the classical latent class model with conditional independence, would be represented by white cells off the diagonal and black cells on the diagonal.

This approach is very general, since any distribution can be chosen for each block as soon as it is different from the distribution of independence. The mixture model by conditional independent blocks is a parsimonious version of the log-linear mixture model. Indeed, the distribution of variables in blocks determines which interactions are modeled. The interactions between variables of different blocks will be zero and those between variables of the same block can be modeled by the specific distribution of the block. The limiting case of this model where $\textsc{b}_k=d$ for each class is equivalent to the latent class model with the conditional independence assumption.

The generic identifiability of the mixture model with conditionally independent blocks follows, under specific constraints, from Theorem 4 of \citet{All09} by assuming that the distribution of each block is itself identifiable. This proof is given in Appendix A of \citet{Mar13}.

\section{Intra-block parsimonious distribution\label{ourmodel}}

The goal is now to define a parsimonious distribution for each block that takes into account the dependency between variables. Furthermore, the parameters of the distribution inside the block must be meaningful for the practitioner. In this context, we propose to model the distribution of each block by a mixture of the extreme distributions according to the Cramer’s V criterion computed on all the variable couples. The model results in a bi-component mixture between an independence distribution and a maximum dependency distribution which can be easily interpreted by the user. The maximum dependency distribution is introduced first. The resulting conditional correlated model (CCM) is also defined as a block model extension of the latent class model where the distribution inside the block is modeled by this bi-component mixture.
Remark: Without loss of generality, the variables are considered as ordered by decreasing number of modalities in each block: ∀(k, b) m{kb}
j ≥ m{kb}
j

The goal is now to define a parsimonious distribution for each block that takes into account the dependency between variables. Furthermore, the parameters of the distribution inside the block must be meaningful for the practitioner. In this context, we propose to model the distribution of each block by a mixture of the extreme distributions according to the Cramer’s V criterion computed on all the variable couples. The model results in a bi-component mixture between an independence distribution and a maximum dependency distribution which can be easily interpreted by the user. The maximum dependency distribution is introduced first. The resulting conditional correlated model (\textsc{ccm}) is also defined as a block model extension of the latent class model where the distribution inside the block is modeled by this bi-component mixture.\\
\textbf{Remark: }Without loss of generality,  the variables are considered as ordered by decreasing number of modalities in each block:  $\forall(k,b)\; m_j^{\{kb\}}\geq m_{j+1}^{\{kb\}}$ where $j=1,\ldots,d^{\{kb\}}-1$.

\subsection{Maximum dependency distribution}
The maximum dependency distribution is defined as the ``opposite'' distribution of independence according to the Cramer's V criterion computed on all the variable couples. Indeed, this latter minimizes this criterion while the maximum dependency distribution maximizes it. Under this distribution, the modality knowledge of one variable provides the maximum information on all the subsequent variables. Note that it is a non-reciprocal functional link between variables. Indeed, if $\boldsymbol{x}^{\{kb\}}$ arises from this distribution, the knowledge of the variable having the largest number of modalities determines exactly the others but the reverse does not necessarily apply. So, this distribution defines successive surjections from  the space of $x^{\{kb\} j}$ to the space of $x^{\{ kb \} j+1}$ with $j=1,\ldots,d^{\{kb\}}-1$ (recall that the variables are ordered by decreasing number of modalities in each block). In fact, it is a reciprocal functional link only when $m_j^{\{kb\}}=m_{j+1}^{\{kb\}}$. 

Since the first variable determines the other ones, this distribution is defined by a product between the multinomial distribution of the first variable parametrized by $\boldsymbol{\tau}_{kb}=(\tau_{kb}^{h};h=1,\ldots,m_1^{\{kb\}})$ with $\tau_{kb}^h\geq 0$ and $\sum_{h=1}^{m^{\{kb\}}_1}\tau_{kb}^h =1$, and the product between the conditional distributions defined as specific multinomial distributions. So, conditionally on $x^{\{kb\}1h}=1$, $\boldsymbol{x}^{\{kb\}j}$ is deterministic for each $j=2,\ldots,d^{\{kb\}}$. Indeed, in such a case, $\boldsymbol{x}^{\{kb\}j}$ follows a specific multinomial distribution whose parameters are 0 and 1. More precisely, this distribution is parametrized by $\boldsymbol{\delta}_{kb}^{hj}=(\delta_{kb}^{hjh'};h'=1,\ldots,m_j^{\{kb\}})$ with the following constraints defining the successive surjections: $\delta_{kb}^{hjh'} \in \{0,1\}$,  $\sum_{h'=1}^{m_j^{\{kb\}}} \delta_{kb}^{hjh'}=1$ (multinomial distribution) and $\sum_{h=1}^{m_1^{\{kb\}}} \delta_{kb}^{hjh'}\geq 1$ (surjections).

By denoting $\boldsymbol{\delta}_{kb}=(\boldsymbol{\delta}_{kb}^{hj};h=1,\ldots,m_1^{\{kb\}};j=2,\ldots,d^{\{kb\}})$, the pdf of maximum dependency distribution is defined as:
\begin{align}
\acute{p}(\boldsymbol{x}^{\{kb\}};\boldsymbol{\tau}_{kb},\boldsymbol{\delta}_{kb})&=
  p(\boldsymbol{x}^{\{kb\}1};\boldsymbol{\tau}_{kb}) \prod_{j=2}^{d^{\{kb\}}} p(\boldsymbol{x}^{\{kb\}j}|\boldsymbol{x}^{\{kb\}1} ;\{\boldsymbol{\delta}_{kb}^{hj}\}_{h=1,\ldots ,m^{\{kb\}}_1})  \nonumber \\ 
&=\prod_{h=1}^{m_1^{\{kb\}}} \Big( \tau_{kb}^h \prod_{j=2}^{d^{\{kb\}}}
\prod_{h'=1}^{m_{j}^{\{kb\}}}  (\delta_{kb}^{hjh'} )^{x^{\{kb\}jh'} } \Big)^{x^{\{kb\}1h}} .
\end{align}

\begin{figure}[h!]
\begin{center}
\begin{minipage}{0.44\textwidth}
\centering \centering \includegraphics[scale=0.35]{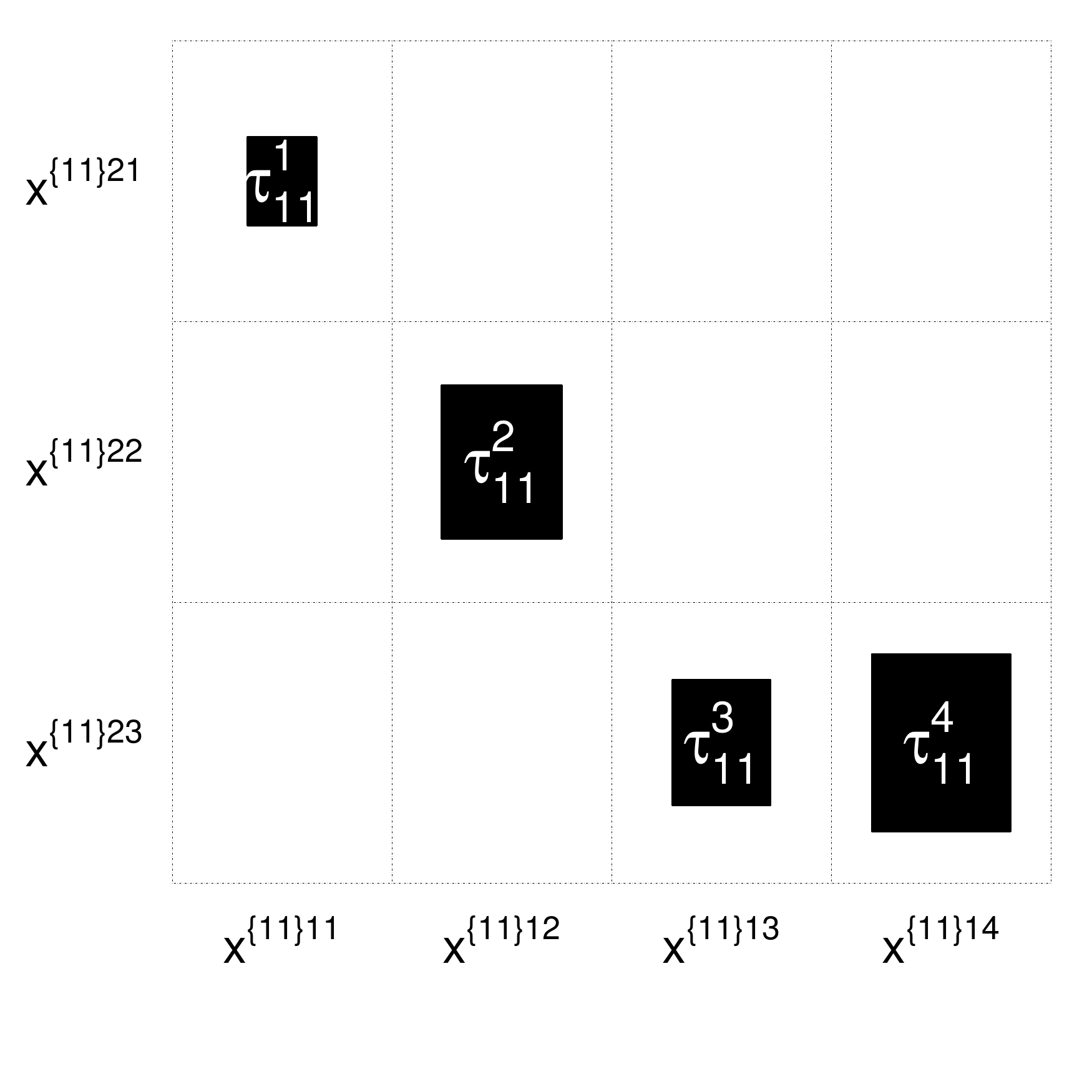} \\
\centering (a)

\end{minipage} 
 \hfill
\begin{minipage}{0.55\textwidth}
\centering \centering \includegraphics[scale=0.4]{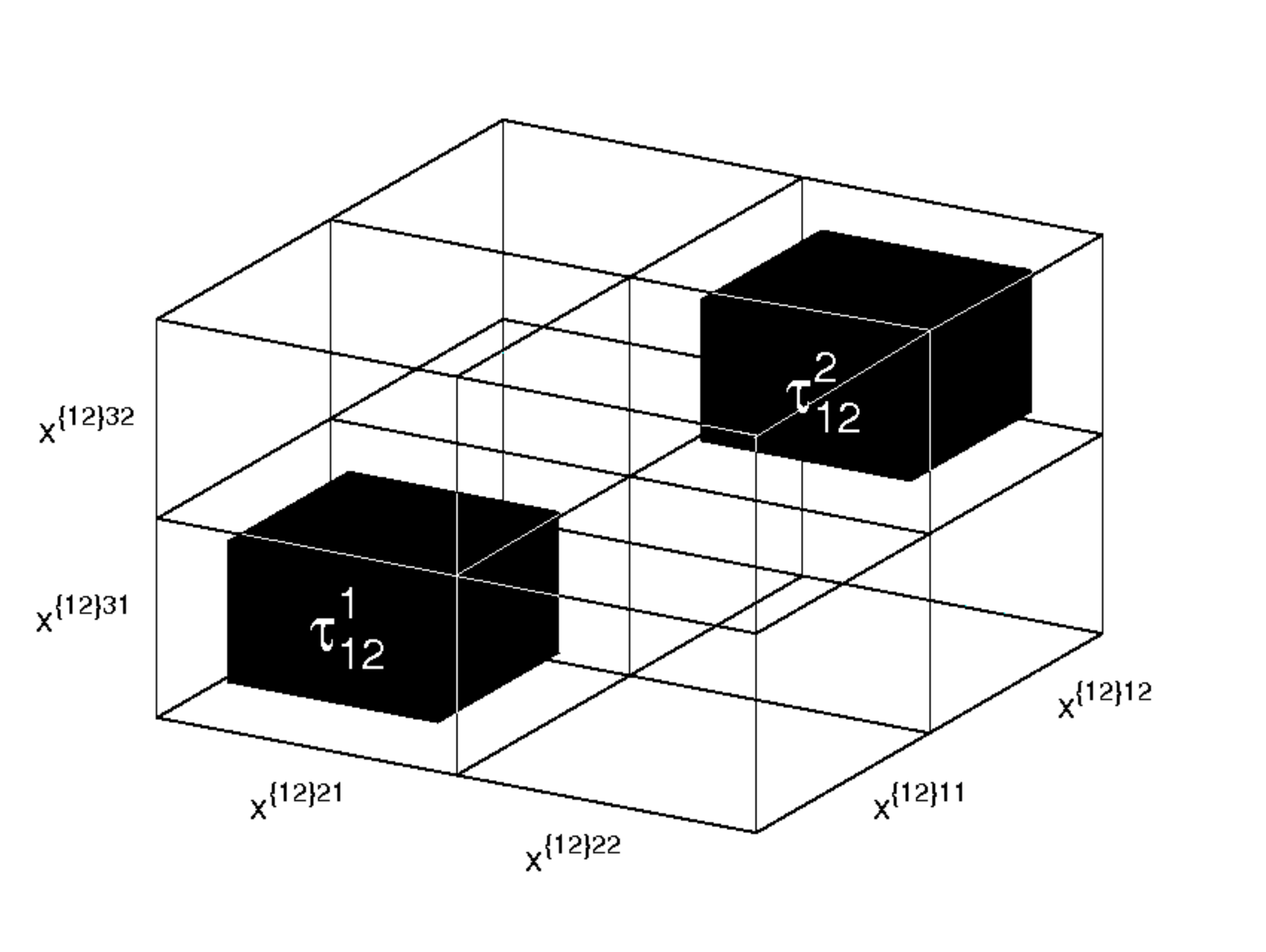}  \\
\centering (b)
\end{minipage} 
 \end{center}
 \centering \caption{Two examples of the maximum dependency distributions for the first component of the mixture illustrated by Figure~\ref{exeblo}(a). (a) The first block is displayed with $m^{\{11\}1}=4$, $m^{\{11\}2}=3$, $\delta^{h1h}_{11}=1$ for $h=1,2,3$, $\delta^{413}_{11}=1$ and $\boldsymbol{\tau}_{11}=(0.1,0.3,0.2,0.4)$; (b) The second block  is displayed with  $m^{\{12\}1}=m^{\{12\}2}=m^{\{12\}3}=2$, $\delta^{hjh'}_{12}=1$ iff ($h=h'$) and $\boldsymbol{\tau}_{12}=(0.5,0.5)$.}\label{exem}
 \end{figure}

Figure~\ref{exem} shows two examples of the maximum dependency distributions. The probabilities of the joint distribution are represented by the area of dark boxes. Notice that $\boldsymbol{\delta}_{kb}$ defines the position where the probabilities are non-zero (location of a dark boxes) and $\boldsymbol{\tau}_{kb}$ defines the probabilities of this non-zero cells (area of the dark boxes). For the example illustrated in Figure~\ref{exem}.a, the probability that the first variable takes the modality one is 0.1 ($\tau_{11}^1=0.1$). Moreover, conditionally on $x^{\{11\}11}=1$, variable $\boldsymbol{x}^{\{11\}2}$ is deterministic since the probability that $x^{\{11\}21}=1$ given $x^{\{11\}11}=1$ is one.

A sufficient condition of identifiability is to impose $\tau^{h}_{kb}>0$ for all $h=1,\ldots,m_1^{\{kb\}}$. This distribution has very limited interest because it is so unrealistic that it can almost never be used alone. We will see in the next section how to use it in a more efficient way.

\subsection{A new block distribution: mixture of two extreme distributions\label{blockdistribution}}
We proposed to model the distribution of each block by a bi-components mixture between an \emph{independence} distribution and a \emph{maximum dependency} distribution. For block $b$ of component $k$, the block distribution is modeled by:
\begin{equation}
p(\textbf{\textit{x}}^{\{kb\}};\boldsymbol{\theta}_{kb})=(1-\rho_{kb})\mathring{p}(\textbf{\textit{x}}^{\{kb\}};\boldsymbol{\alpha}_{kb})+\rho_{kb}\acute{p}(\textbf{\textit{x}}^{\{kb\}};\boldsymbol{\tau}_{kb},\boldsymbol{\delta}_{kb}),\label{DDI}
\end{equation} 
where  $\boldsymbol{\theta}_{kb}=(\rho_{kb},\boldsymbol{\alpha}_{kb},\boldsymbol{\tau}_{kb},\boldsymbol{\delta}_{kb})$ and where $\rho_{kb}$ is the proportion of the maximum dependency distribution in this mixture with $0\leq\rho_{kb} \leq 1$. The proposed model requires few additional parameters compared with the conditional independence model. In addition, it is easily interpretable as explained in the next paragraph. Note that the limiting case where $\rho_{kb}=0$ considers that the block follows an independence distribution. In this particular case, the parameters of the maximum dependency distribution are no longer defined.

Under this distribution, the proportion of the maximum dependency distribution reflects the deviation from independence  under the assumption that the other allowed distribution is the maximum dependency distribution. The parameter \emph{$\rho_{kb}$}  gives an indicator of the \emph{inter-variables correlation} of the block. It is not here a pairwise dependency among variables but a dependency between all variables of the block. Furthermore, it stays bounded when the number of variables is larger than two while the Cramer's V is non-upper-bounded in this case. The \emph{intra-variables dependencies} between the variables are defined by $\boldsymbol{\delta}_{kb}$. The strength of these dependencies is explained by $\boldsymbol{\tau}_{kb}$ since it gives the \emph{weight of the over-represented modalities crossing } compared with the independence distribution.

Above, we interpreted the distribution by conditionally independent blocks as a parsimonious version of the  log-linear mixture model because it determines the interactions to be modeled for each class. By choosing the proposed distribution for blocks, a second level of parsimony is added. Indeed, among the interactions allowed by this distribution with independent blocks, only those corresponding to the maximum dependency distribution will be modeled. Other interactions are considered as null. 
 
\paragraph*{Properties:}
\begin{itemize}
\item The \textsc{ccm}, stays parsimonious compared with \textsc{cim} since, for each block with at least two variables, the number of the additional parameters depends only on the number of modalities of the first variable of the block and not on the number of variables in the block. By using $\nu_{\textsc{cim}}$ defined in Equation~\eqref{nbparamcim}, the number of parameters of \textsc{ccm} is denoted $\nu_{\textsc{ccm}}$ by:
\begin{equation}
\nu_{\textsc{ccm}}=\nu_{\textsc{cim}}+\sum_{\{(k,b)|d^{\{kb\}}>1\}}m_1^{\{kb\}}.
\end{equation}
\item The proposed distribution is identifiable under the condition that the block is composed by at least three variables ($d^{\{kb\}}>2$) or that the modality number of the last variable of the block is more than two ($m_2^{\{kb\}}>2$). This result is demonstrated in Appendix B of \citet{Mar13}. The parameter $\rho_{kb}$ is a new indicator allowing measuring the correlation between variables, not limited to correlation between variable couples.
In the case where the identifiability conditions cannot be met, we distinguish two cases. If $d^{\{kb\}}=1$, then block $b$ contains only one variable, and the proposed model is reduced to model a multinomial distribution, $\rho_{kb}=0$ and the maximum dependency distribution is not defined. If $d^{\{kb\}}=2$ and $m_2^{\{kb\}}=2$ then a new constraint is added. In order to have the most meaningful parameters, the chosen value of $\rho_{kb}$ is the largest value maximizing the log-likelihood. This additional constraint does not falsify the definition of $\rho_{kb}$ as an indicator of the dependency strength between the variables of the same block. Furthermore, this constraint is natural since blocks with the biggest dependencies are wanted. Note that $\rho_{kb}$ seems to be correlated with the Cramer's V. An example is given in Section 3 of \citet{Mar13}.
\item Note that the marginal probabilities given class can be straightforwardly deduced from the parameters of the model.
\end{itemize}

\section{Estimation of the parameters \label{estimation}}
For a fixed model $(g,\boldsymbol{\sigma})$, the parameters must be estimated. Since the proposed distribution \textsc{ccm} has two latent variables (the class membership and the intra-block distribution membership), two algorithms derived from the EM algorithm perform the estimation of the associated continuous parameters. The combinatorial problems arising from the consideration of the discrete parameters are avoided by using a Metropolis-Hastings algorithm.

\subsection{Global GEM algorithm \label{EMglo}}
The whole data set consisting of $n$ independent and identically distributed individuals is denoted by $\textbf{x} = (\textbf{\textit{x}}_{1},\ldots\textbf{\textit{x}}_{n} )$ where $\textbf{\textit{x}}_{i} \in \mathcal{X}$. The objective is to obtain the maximum log-likelihood estimator $\hat{\boldsymbol{\theta}}$  defined as ($g$ is now implicit in each expression)
	\begin{equation}
\hat{\boldsymbol{\theta}}=\text{argmax}_{\boldsymbol{\theta}} L(\boldsymbol{\theta} ; \mathbf{x},\boldsymbol{\sigma}) \quad \text{with} \quad	L(\boldsymbol{\theta} ; \mathbf{x},\boldsymbol{\sigma})=\sum_{i=1}^{n} \ln \Big(  p(\textbf{\textit{x}}_{i};\boldsymbol{\sigma},\boldsymbol{\theta}) \Big). \label{vrai}
	\end{equation}

The search for maximum likelihood estimates for mixture models entails solving equations having no analytical solutions. For the mixture models, the assignments of the individuals into the classes can be considered as missing data. So, the tool generally used is the Expectation-Maximization algorithm (denoted EM algorithm) and its extensions \citep*{Dem77,Mcl97}. We denote the unknown indicator vectors of the $g$ clusters by $\textbf{z} = (z_{ik};i=1,\ldots,n;k=1,\ldots,g)$ where $z_{ik} = 1$ if $\textbf{\textit{x}}_{i}$ arises from cluster $k$, $z_{ik} = 0$ otherwise. Thus, the mixture model distribution corresponds to the marginal distribution of the random variable $\textbf{X}$ obtained from the couple distribution of the random variables $(\textbf{X},\textbf{Z})$. In order to maximize the log-likelihood, the EM algorithm uses the complete-data log-likelihood which is defined as
	\begin{equation}
	L_{\text{c}}(\boldsymbol{\theta} ; \mathbf{x},\textbf{z},\boldsymbol{\sigma})=\sum_{i=1}^{n}\sum_{k=1}^{g} z_{ik} \ln \Big( \pi_{k} p(\textbf{\textit{x}}_{i};\boldsymbol{\sigma}_k,\boldsymbol{\theta}_{k}) \Big) . \label{vraic}
	\end{equation}
	
The EM algorithm is an iterative algorithm which alternates between two steps: the computation of the complete-data log-likelihood conditional expectation (E step) and its maximization (M step). Many algorithms are derived from the EM algorithm and among them the Generalized EM algorithm (GEM) is of interest to us. It works on the same principle as the EM algorithm, but the maximization step is replaced by a GM step where the proposed parameters increase the expectation of the complete-data log-likelihood according to its previous value without necessarily maximizing it. 

We prefer to use the GEM algorithm, since the maximization step in the EM algorithm requires  estimating the continuous parameters for too many possible values of the discrete parameters in order to warrant the maximization of the complete-data log-likelihood expectation. Indeed, exhaustive enumeration for estimating the discrete parameters is generally impossible when a block contains variables with many modalities and/or many variables, as detailed below. If $S(a,b)$ is the number of possible surjections from a set of cardinal $a$ into a set of cardinal $b$, then $\boldsymbol{\delta}_{kb}$ is defined in the discrete space of dimension $\prod_{j=1}^{d^{\{kb\}}-1} S(m_j^{\{kb\}},m_{j+1}^{\{kb\}})$. For example, a block with three variables and $m^{\{kb\}}=(5,4,3)$ implies $51,840$ possibilities for $\boldsymbol{\delta}_{kb}$. Thus, a stochastic approach is proposed in Section~\ref{discestim} to overcome this problem. Then, the estimation of the continuous parameters conditionally on the discrete parameters is performed via the classical EM algorithm presented in Section~\ref{continuestim} since their estimation cannot be obtained in closed form. At  iteration  $(r)$, the steps of the global GEM can be written as:
\begin{itemize}
\item \textbf{$\text{E}_{\text{global}}$ step:} 
${z}_{ik}^{(r)}=\frac{\pi_k^{(r)}p(\boldsymbol{x}_i;\boldsymbol{\sigma}_k,\boldsymbol{\theta}_k^{(r)})}{\sum_{k'=1}^g \pi_{k'}^{(r)}p(\boldsymbol{x}_i;\boldsymbol{\sigma}_{k'},\boldsymbol{\theta}_{k'}^{(r)})}
$,
\item  \textbf{$\text{GM}_{\text{global}}$ step:} 
 $\pi^{(r+1)}_{k}=\frac{n_k^{(r)}}{n}$ where $n_k^{(r)}=\sum_{i=1}^n {z}_{ik}^{(r)}$ and $\forall (k,b)$ $\boldsymbol{\theta}_{kb}^{(r+1)}$ is updated under the constraint that the conditional expectation of complete-data log-likelihood increases (see Sections \ref{discestim} and \ref{continuestim}).
\end{itemize}

\paragraph*{Initialization of the algorithm:} Since this algorithm is performed  in an stochastic algorithm used for the model selection (see Section \ref{modelsel}) and since this latter has an influence on the GEM initialization, this point will be detailed in Section~\ref{MCMCapprox}.

\paragraph*{Stopping criterion:} The GEM algorithm is stopped after $r_{\max}$ iterations and we fix $\hat{\boldsymbol{\theta}}=\boldsymbol{\theta}^{(r_{\max})}$.

\subsection{Details of the $\text{GM}_{\text{global}}$ step of the GEM\label{discestim}}
\begin{figure}[h!]
\begin{minipage}{0.19\textwidth}
\centering \includegraphics[scale=0.18]{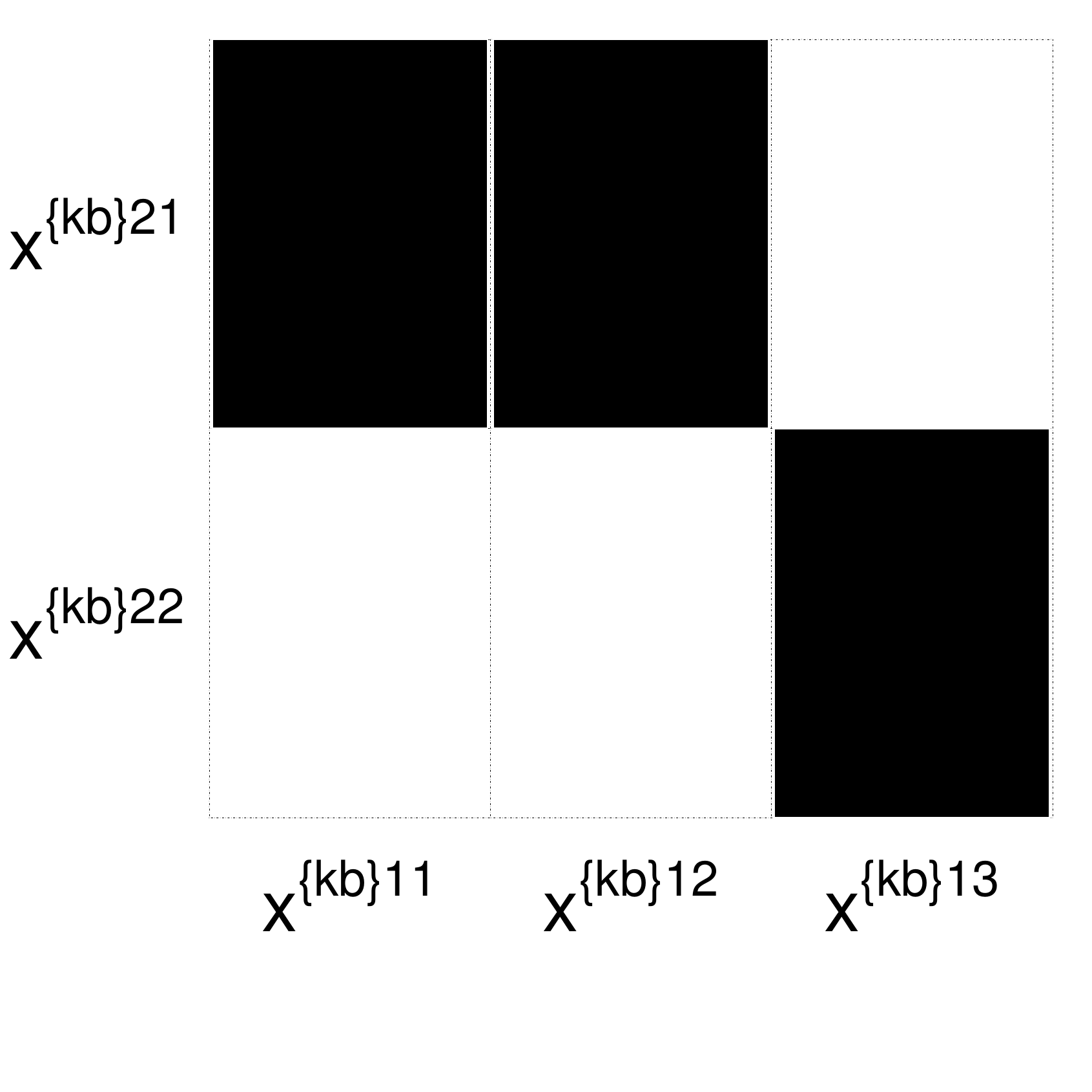} \\
\centering (a)
\end{minipage}
\begin{minipage}{0.19\textwidth}
\centering \includegraphics[scale=0.18]{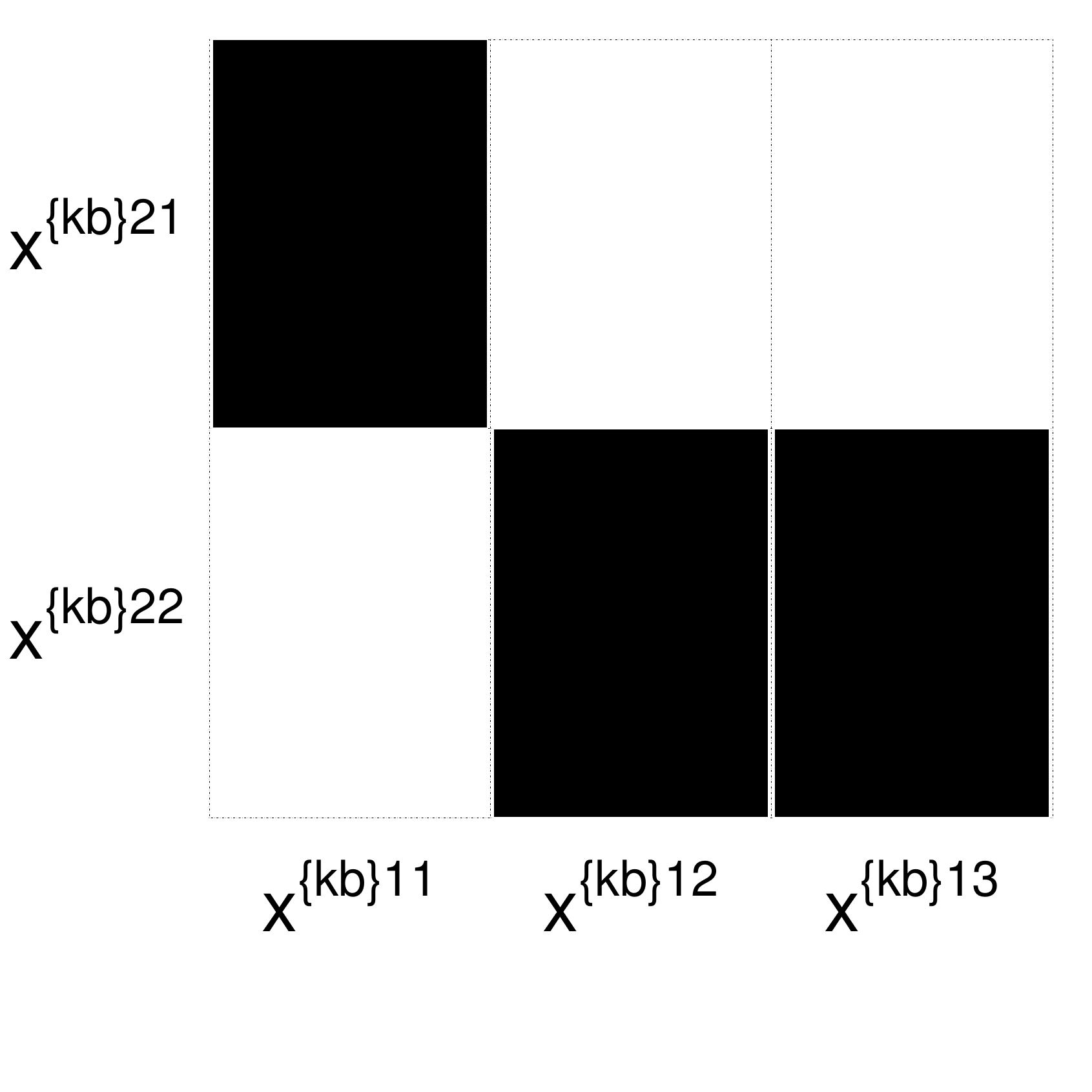} \\
\centering (b.1)
\end{minipage}
\begin{minipage}{0.19\textwidth}
\centering \includegraphics[scale=0.18]{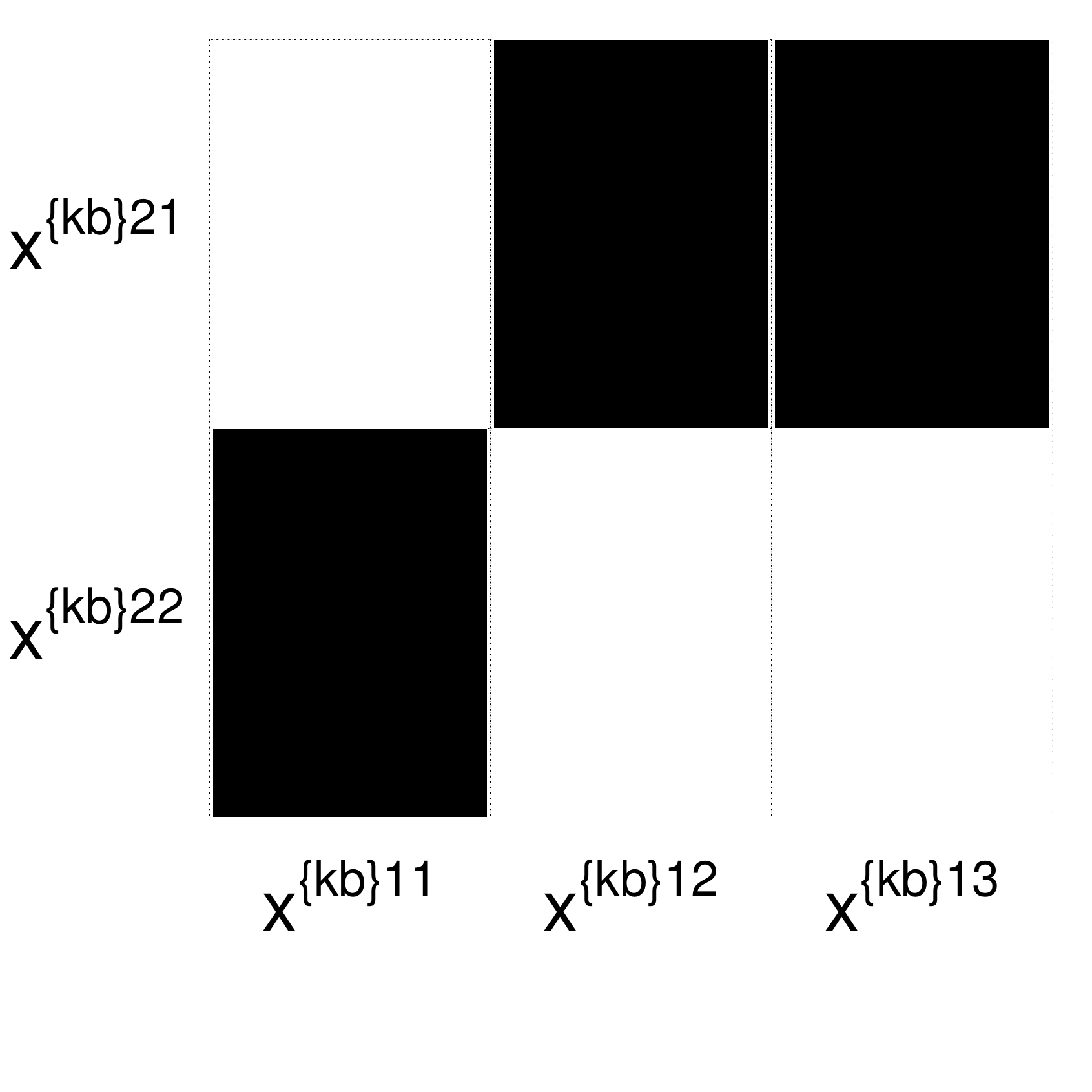}\\
\centering (b.2)
\end{minipage}
\begin{minipage}{0.19\textwidth}
\centering \includegraphics[scale=0.18]{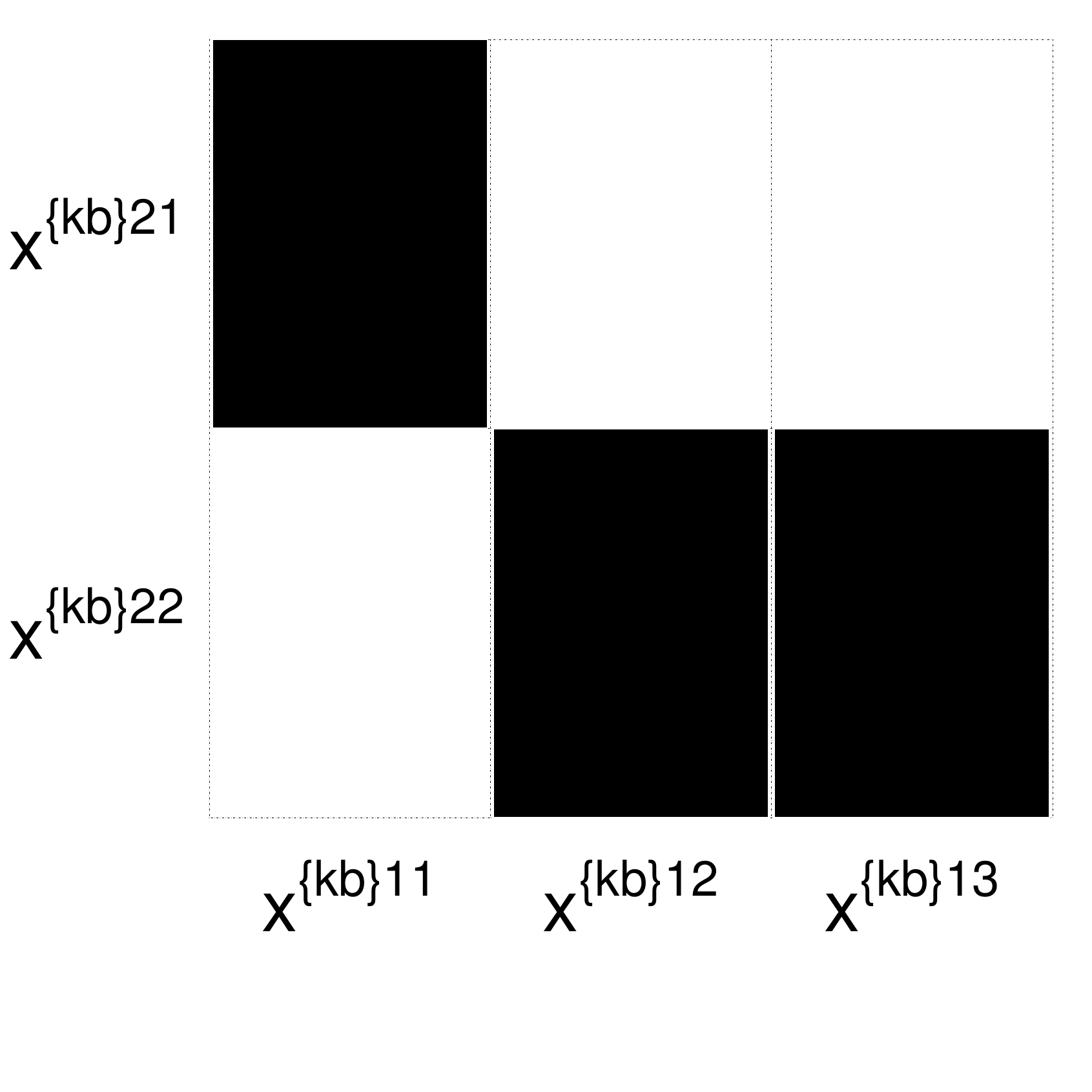} \\
\centering (b.3)
\end{minipage}
\begin{minipage}{0.19\textwidth}
\centering \includegraphics[scale=0.18]{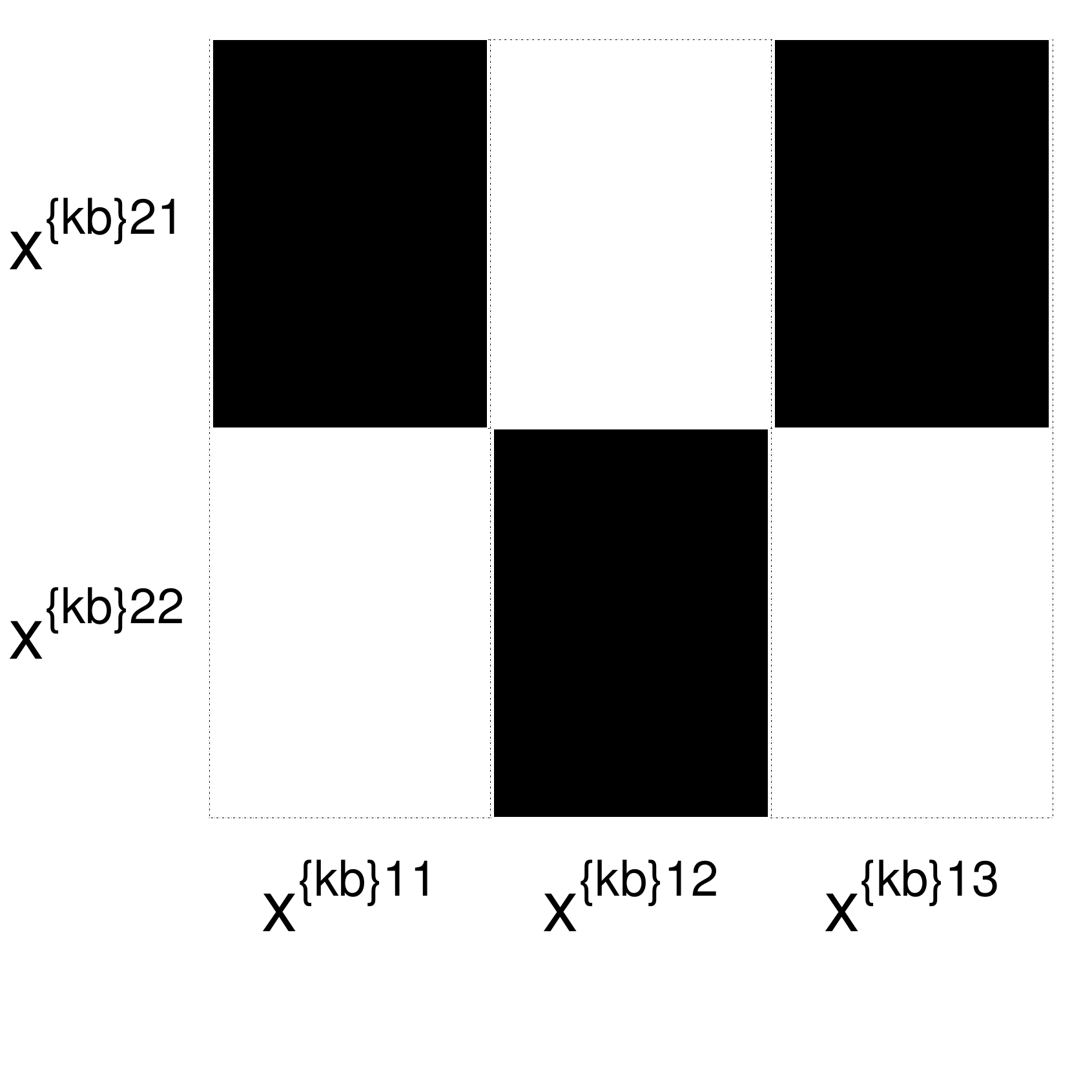}\\
\centering (b.4)
\end{minipage}
\caption{Example of $\Delta(\boldsymbol{\delta}_{kb})$ with $d^{\{kb\}}=2$ and $\boldsymbol{m}^{\{kb\}}=(3,2)$. For row $h'$ and column $h$, a black cell indicates that $\delta_{kb}^{h2h'}=1$ and a white cell that $\delta_{kb}^{h2h'}=0$: (a) $\boldsymbol{\delta}_{kb}$; (b.1), (b.2), (b.3), (b.4) are the elements of $\Delta(\boldsymbol{\delta}_{kb})$. \label{voisindelta}}
\end{figure}

The maximization of the expected complete-data log-likelihood is done by optimizing its terms for each $(k,b)$. Thus, the determination of $\boldsymbol{\theta}_{kb}^{(r+1)}$ is performed independently of the parameters of the other blocks. A Metropolis-Hastings algorithm \citep{Rob04} is also performed, for each $(k,b)$, to avoid the combinatorial problems induced by the detection of the discrete parameters $\boldsymbol{\delta}_{kb}$. It performs a random walk over the discrete parameters space and computes the maximum likelihood estimators of the continuous parameters $(\rho_{kb},\boldsymbol{\alpha}_{kb},\boldsymbol{\tau}_{kb})$ associated with them. This stochastic algorithm allows finding the estimator maximizing the expected complete-data log-likelihood of block $b$ for component $k$:
\begin{equation}
\underset{\boldsymbol{\theta}_{kb}}{\text{argmax }} \sum_{i=1}^n z_{ik}^{(r)} \ln p(\boldsymbol{x}^{\{kb\}}_i;\boldsymbol{\theta}_{kb}).
\end{equation}
At each iteration $(s)$ of this Metropolis-Hastings algorithm, a discrete parameter denoted by $\boldsymbol{\delta}_{kb}^{(r,s+\frac{1}{2})} $ is sampled with a uniform distribution in a neighborhood of $\boldsymbol{\delta}_{kb}^{(r,s)}$ denoted as $\Delta(\boldsymbol{\delta}_{kb}^{(r,s)})$. Then the continuous parameters $(\rho_{kb}^{(r,s+\frac{1}{2})},\boldsymbol{\alpha}_{kb}^{(r,s+\frac{1}{2})},$ $\boldsymbol{\tau}_{kb}^{(r,s+\frac{1}{2})})$  are computed, conditionally on the value of  $\boldsymbol{\delta}_{kb}^{(r,s+\frac{1}{2})}$, in order to maximize the expected complete-data log-likelihood of block $b$ for component $k$:
\begin{equation}
\sum_{i=1}^n z_{ik}^{(r)} \ln  p(\boldsymbol{x}^{\{kb\}}_i;\rho_{kb},\boldsymbol{\alpha}_{kb},\boldsymbol{\tau}_{kb},\boldsymbol{\delta}_{kb}^{(r,s+\frac{1}{2})} ).
\end{equation}
The candidate parameters are now denoted by $\boldsymbol{\theta}_{kb}^{(r,s+\frac{1}{2})} =(\rho_{kb}^{(r,s+\frac{1}{2})} ,\boldsymbol{\alpha}_{kb}^{(r,s+\frac{1}{2})} ,\boldsymbol{\tau}_{kb}^{(r,s+\frac{1}{2})} ,\boldsymbol{\delta}_{kb}^{(r,s+\frac{1}{2})} )$. The whole block parameters $\boldsymbol{\theta}_{kb}^{(r,s+1)}$ of the next step are then defined as $\boldsymbol{\theta}_{kb}^{(r,s+\frac{1}{2})} $ with the acceptance probability $\mu^{(r,s+1)}$ and $\boldsymbol{\theta}_{kb}^{(r,s)}$ otherwise, where: 

\begin{equation}
\mu^{(r,s+1)}=\min\left\{\frac{ \prod_{i=1}^n p(\boldsymbol{x}^{\{kb\}}_i;\boldsymbol{\theta}_{kb}^{(r,s+\frac{1}{2})} )^{z_{ik}^{(r)}}|\Delta(\boldsymbol{\delta}_{kb}^{(r,s+\frac{1}{2})} )|}{ \prod_{i=1}^n p(\boldsymbol{x}^{\{kb\}}_i;\boldsymbol{\theta}_{kb}^{(r,s)})^{z_{ik}^{(r)}}|\Delta(\boldsymbol{\delta}_{kb}^{(r,s)})|
},1\right\},
\end{equation}
$|\Delta(\boldsymbol{\delta}_{kb}^{(r,s)})|$ denoting the cardinal of $\Delta(\boldsymbol{\delta}_{kb}^{(r,s)}) $. Thus, at iteration $(s)$, the algorithm performs the three following steps:
\begin{itemize}
\item \textbf{Stochastic step on $\boldsymbol{\delta}_{kb}$:} generate $\boldsymbol{\delta}_{kb}^{(r,s+\frac{1}{2})}$  with a uniform distribution among the elements of $\Delta(\boldsymbol{\delta}_{kb}^{(r,s)})$,
\item \textbf{Maximization step on the continuous parameters ($\text{M}_{\boldsymbol{\theta}}$ step):} compute the continuous parameters of $\boldsymbol{\theta}_{kb}^{(r,s+\frac{1}{2})}$ (see Section~\ref{continuestim}),
\item \textbf{Stochastic step on $\boldsymbol{\theta}_{kb}$:}  sample $
\boldsymbol{\theta}_{kb}^{(r,s+1)}=\left\{
\begin{array}{rl}
\boldsymbol{\theta}_{kb}^{(r,s+\frac{1}{2})} & \text{ with probability } \mu^{(r,s+1)}
 \\
\boldsymbol{\theta}_{kb}^{(r,s)} & \text{ otherwise. }
\end{array}\right. 
$	
\end{itemize}
Neighborhood  $\Delta(\boldsymbol{\delta}_{kb}^{(r,s)})$ is defined as the set of parameters where at most two surjections are different from that of $\boldsymbol{\delta}_{kb}^{(r,s)}$. Figure~\ref{voisindelta} illustrates this definition.

\paragraph*{Initialization of the algorithm:}
The initialization of the  algorithm is done by $\boldsymbol{\theta}_{kb}^{(r+1,0)}=\boldsymbol{\theta}_{kb}^{(r)}$. 
 
\paragraph*{Stopping criterion:}
This algorithm is stopped after a number of iterations $s_{\max}$. The parameter $\boldsymbol{\theta}_{kb}^{(r+1)}=\boldsymbol{\theta}_{kb}^{(r+1,\tilde{s})}$ is returned with $\tilde{s}=\underset{s}{\text{argmax}}  \sum_{i=1}^n z_{ik}^{(r)} \ln p(\boldsymbol{x}^{\{kb\}}_i;\boldsymbol{\theta}_{kb}^{(r,s)})$. Thus, the proposed initialization ensures an increased likelihood at each iteration of the GEM algorithm.

\paragraph*{Remark: }
When the space of possible $\boldsymbol{\delta}_{kb}$ is small (for example when the block groups a small number of binary variables), an exhaustive approach obtains the same results as the proposed algorithm with less computation time. Thus, the retained approach (exhaustive or stochastic) depends on the number of variables and modalities.

\subsection{Details of the $\text{M}_{\boldsymbol{\theta}}$ step of the $\text{GM}_{\text{global}}$ step\label{continuestim}}

As there is a second level of mixing, another EM algorithm can be performed for the continuous parameters $(\rho_{kb},\boldsymbol{\alpha}_{kb},\boldsymbol{\tau}_{kb})$ estimation by introducing other unknown vectors corresponding to the indicator of the blocks distributions conditionally on $\textbf{z}$. These vectors are denoted by $\textbf{y}=(\textbf{y}^{\{kb\}};k=1,\ldots,g;b=1,\ldots,\textsc{b}_k)$ with $\textbf{y}^{\{kb\}}=(y_{1}^{\{kb\}},...,y_{n}^{\{kb\}})$ where $y_{i}^{\{kb\}}=1$ if $\textbf{\textit{x}}_{i}^{\{kb\}}$ arises from the \emph{maximum dependency} distribution for  block $b$ of cluster $k$ and $y_{i}^{\{kb\}}=0$ if $\textbf{\textit{x}}_{i}^{\{kb\}}$ arises from the \emph{independence} distribution for block $b$ of cluster $k$. The whole mixture model distribution corresponds to the marginal distribution of the random variable $\textbf{X}$ obtained from the triplet distribution of the random variables $(\textbf{X},\textbf{Y},\textbf{Z})$. Since the blocks are independent conditionally on $\textbf{Z}$, the \emph{full} complete-data log-likelihood (both in $\textbf{Y}$ and $\textbf{Z}$) is defined as:
\begin{small}
\begin{align}
L_{\text{c}}^{\text{full}}(\boldsymbol{\theta}; \mathbf{x},\textbf{y},\textbf{z},\boldsymbol{\sigma})=\sum_{i=1}^{n}\sum_{k=1}^{g} z_{ik}  \Big( \ln \pi_{k} + \sum_{b=1}^{B_{k}} & \big( (1-y_{i}^{\{kb\}})\ln(1-\rho_{kb}) + (1-y_{i}^{\{kb\}})\ln\mathring{p}(\textbf{\textit{x}}_i^{\{kb\}};\boldsymbol{\alpha}_{kb}) \nonumber \\
 & + y_{i}^{\{kb\}}\ln\rho_{kb} + y_{i}^{\{kb\}}\ln\acute{p}(\textbf{\textit{x}}_i^{\{kb\}};\boldsymbol{\tau}_{kb},\boldsymbol{\delta}_{kb})    \big) \Big).
\end{align}
\end{small}
At iteration $(t)$, the local EM algorithm estimates the continuous parameters of block $b$, with fixed values of  $\textbf{z}^{(r)} $ and  $\boldsymbol{\delta}_{kb}^{(r,s+\frac{1}{2})} $, by the following two steps:
\begin{itemize}
\item \textbf{$\text{E}_{\text{local}}$ step:} 
$y_{i}^{\{kb\}(r,s+\frac{1}{2},t)}=\frac{\rho_{kb}^{(r,s+\frac{1}{2},t)}\acute{p}(\textbf{\textit{x}}_{i}^{\{kb\}};\boldsymbol{\tau}_{kb}^{(r,s+\frac{1}{2},t)},\boldsymbol{\delta}_{kb}^{(r,s+\frac{1}{2})})}{p(\textbf{\textit{x}}^{\{kb\}}_i;\rho_{kb}^{(r,s+\frac{1}{2},t)},\boldsymbol{\alpha}_{kb}^{(r,s+\frac{1}{2},t)},\boldsymbol{\tau}_{kb}^{(r,s+\frac{1}{2},t)},\boldsymbol{\delta}_{kb}^{(r,s+\frac{1}{2})}  )}$,
\item \textbf{$\text{M}_{\text{local}}$  step:} 
$\rho_{kb}^{(r,s+\frac{1}{2},t+1)}=\frac{n_{kb}^{(r,s+\frac{1}{2},t)}}{n_{k}^{(r)}},$   $\tau_{kb}^{(r,s+\frac{1}{2},t+1)}=\frac{\sum_{i=1}^{n}z_{ik}^{(r)}y_{i}^{\{kb\}(r,s+\frac{1}{2},t)} x_{i}^{\{kb\}1h}}{n_{kb}^{(r,s+\frac{1}{2},t)}} $,\\
$\alpha_{kb}^{(r,s+\frac{1}{2},t+1)}=\frac{\sum_{i=1}^{n}z_{ik}^{(r)}(1-y_{i}^{\{kb\}(r,s+\frac{1}{2},t)})x_{i}^{\{kb\}jh}}{n_{k}^{(r)}-n_{kb}^{(r,s+\frac{1}{2},t)}},$ where  $n_{kb}^{(r,s+\frac{1}{2},t)}=\sum_{i=1}^{n}z_{ik}^{(r)}y_{i}^{\{kb\}(r,s+\frac{1}{2},t)}$.
\end{itemize}

\paragraph*{Conjecture:} During our numerous experiments, we empirically noticed that the log-likelihood function of the mixture between the independence and the maximum dependency distributions had a unique optimum. We conjecture that this function has indeed a unique maximum.

\paragraph*{Initialization of the algorithm: } The previous conjecture allows to perform only one initialization of the EM algorithm  fixed to: $(\rho_{kb}^{(r,s+\frac{1}{2},0)},\boldsymbol{\alpha}_{kb}^{(r,s+\frac{1}{2},0)},\boldsymbol{\tau}_{kb}^{(r,s+\frac{1}{2},0)})=(\rho_{kb}^{(r,s)},\boldsymbol{\alpha}_{kb}^{(r,s)},$ $\boldsymbol{\tau}_{kb}^{(r,s)})$.

\paragraph*{Stopping criterion:} This algorithm is stopped after a number of iterations denoted by $t_{\max}$ and returns the value of block parameters $\boldsymbol{\theta}_{kb}^{(r,s+\frac{1}{2})}$ defined as $\boldsymbol{\theta}_{kb}^{(r,s+\frac{1}{2})}=\boldsymbol{\theta}_{kb}^{(r,s+\frac{1}{2},t_{\max})}$.

\paragraph*{Remark: }
In the specific case where $\boldsymbol{\delta}_{kb}$ are known for each $(k,b)$, the estimation of all the continuous parameters could be performed by a unique EM algorithm where, at iteration $(r)$, the E~step would compute both $\textbf{z}^{(r)} $ and $\textbf{y}^{(r)}$ while the M~step would estimate all the parameters maximizing the expectation of the \emph{full} complete-data log-likelihood.

\section{Model selection \label{modelsel} }

\subsection{Gibbs algorithm for exploring the space of models}
Since the number of components $g$ determines the dimension of $\boldsymbol{\sigma}$, the model construction is done in two steps. Firstly, the selection of the number of components and, secondly, the determination of the variable repartition per blocks for each component. In a Bayesian context, the best model $(\hat{g},\hat{\boldsymbol{\sigma}})$ is defined as \citep{Rob05}:
\begin{equation}
(\hat{g},\hat{\boldsymbol{\sigma}})=\underset{g,\boldsymbol{\sigma}}{\text{argmax }}p(g,\boldsymbol{\sigma}|\textbf{x}).
\end{equation}
Thus, by considering that $p(g)=\frac{1}{g_{\max}}$ if $g\leq g_{\max}$ and $0$ otherwise, where $g_{\max}$ is the maximum number of classes allowed by the user, and by assuming that $p(\boldsymbol{\sigma} | g)$ follows a uniform distribution, the best model is also defined as:
\begin{equation}
(\hat{g},\hat{\boldsymbol{\sigma}})=\underset{g}{\text{argmax }} \Big[ \underset{\boldsymbol{\sigma}}{\text{argmax }} p(\textbf{x}|g,\boldsymbol{\sigma}) \Big].
\end{equation}
To obtain $(\hat{g},\hat{\boldsymbol{\sigma}})$, a Gibbs algorithm is used for estimating $ \text{argmax}_{\boldsymbol{\sigma}}\; p(\textbf{x}|g,\boldsymbol{\sigma})$, for each value of $g\in\{1,\ldots,g_{max}\}$. Indeed, this method limits the combinatorial problem involved by the detection of the block structure of variables. A reversible jump method could be used \citep{Ric97}, but this approach is rarely performed with mixed parameters (continuous and discrete). Indeed, in such a case, it is difficult to define a mapping between the parameters space of two models. So, we propose to use an easier Gibbs sampler-type having $p(\boldsymbol{\sigma}|\textbf{x},g)$ as stationary distribution. It alternates between two steps: the generation of a stochastic neighborhood $\Sigma^{[q]}$ conditionally on the current model $\boldsymbol{\sigma}^{[q]}$ by a proposal distribution and the generation of a new pattern $\boldsymbol{\sigma}^{[q+1]}$ included in $\Sigma^{[q]}$ with a probability proportional to its posterior probability. At iteration $[q]$, it is written as:
\begin{itemize}
\item \textbf{Neighborhood step:} generate a stochastic neighborhood $\Sigma^{[q]}$ by a proposal distribution given below conditionally on the current model $\boldsymbol{\sigma}^{[q]}$,
\item  \textbf{Pattern step:} \begin{small}
$\boldsymbol{\sigma}^{[q+1]} \sim p(\boldsymbol{\sigma}|\textbf{x},g, \Sigma^{[q]})$ with $p(\boldsymbol{\sigma}|\textbf{x},g, \Sigma^{[q]})=\left\{\begin{array}{rl}
\frac{p(\textbf{x}|g,\boldsymbol{\sigma})}{\sum_{\boldsymbol{\sigma '}\in  \Sigma^{[q]}}p(\textbf{x}|g,\boldsymbol{\sigma '})} & \text{ if } \boldsymbol{\sigma}\in  \Sigma^{[q]}\\
0 & \text{ otherwise.}	\end{array}\right.$
\end{small}
\end{itemize}

\begin{figure}[h!]
\begin{minipage}{0.16\textwidth}
\centering \includegraphics[scale=0.16]{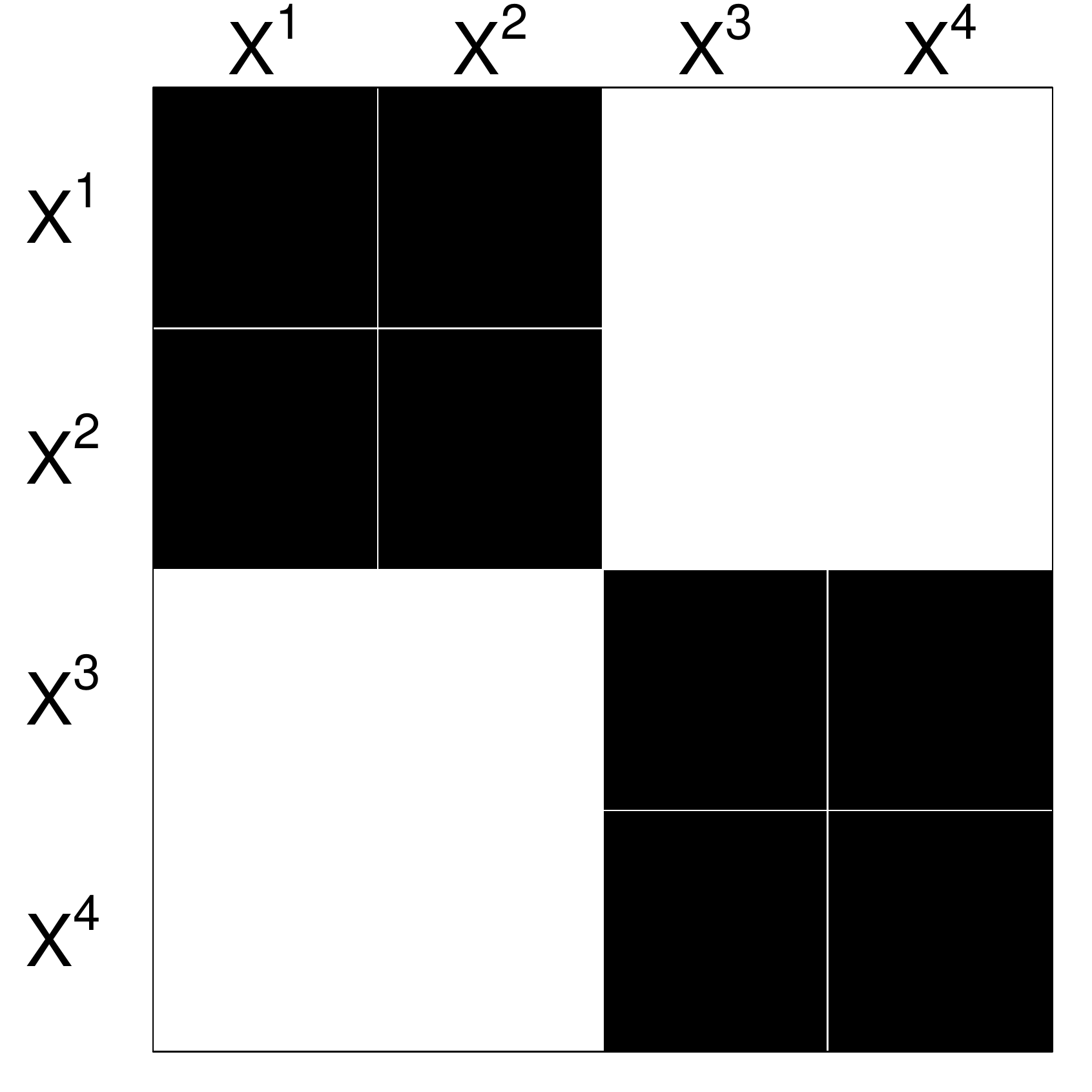} \\
(a)\\
\end{minipage} 
\begin{minipage}{0.89\textwidth}
\centering \includegraphics[scale=0.16]{base_sigma.pdf} 
\centering \includegraphics[scale=0.16]{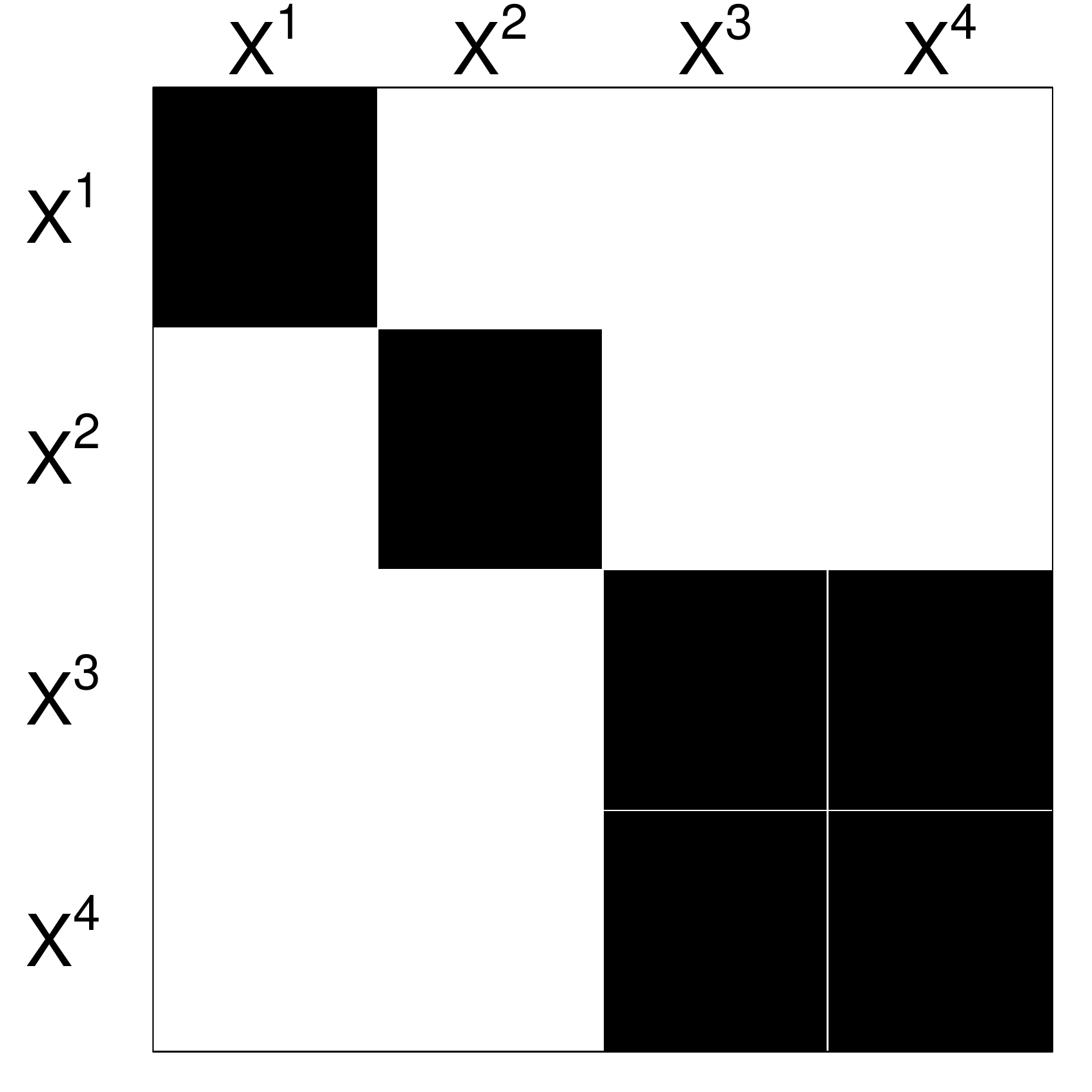}  
\centering \includegraphics[scale=0.16]{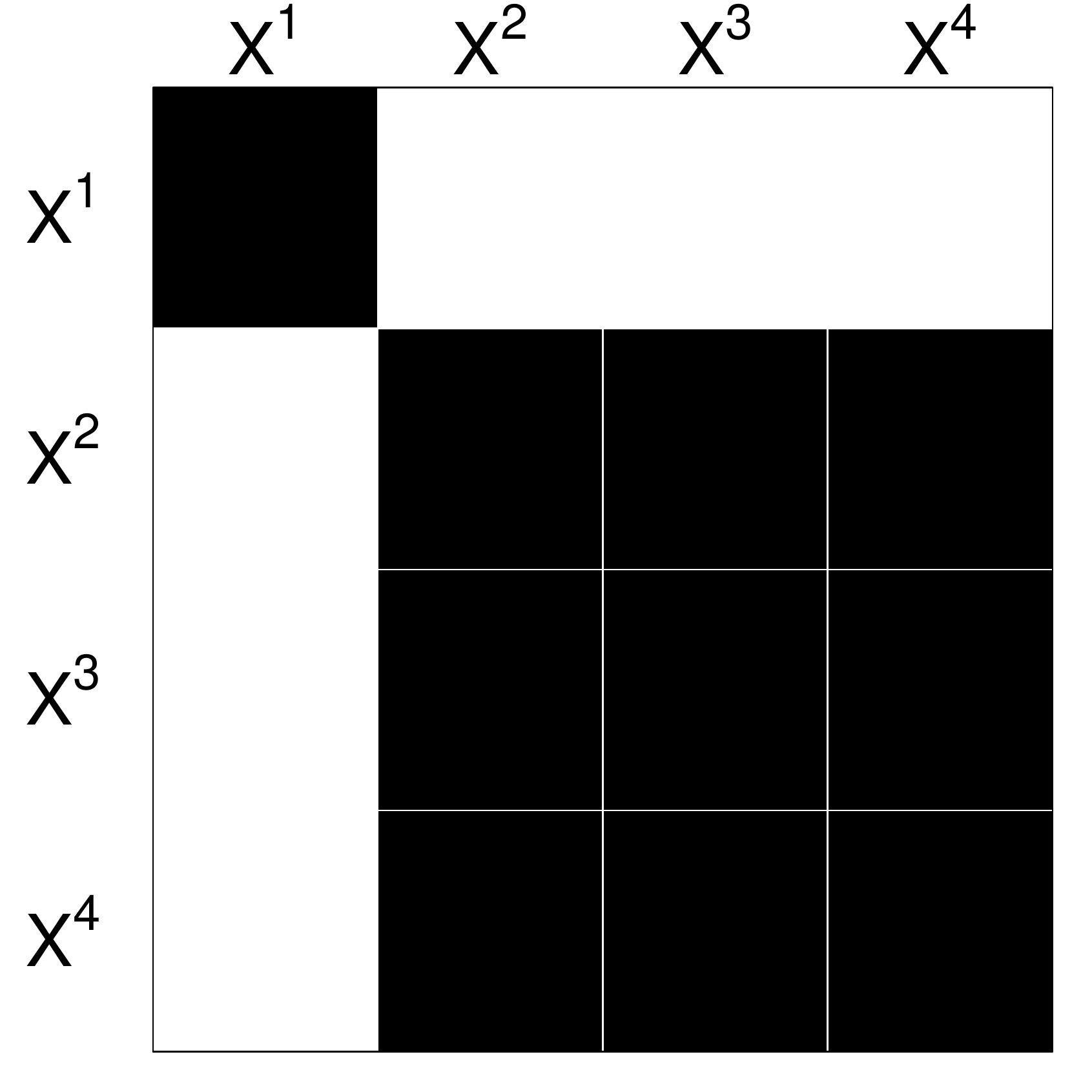} 
\centering \includegraphics[scale=0.16]{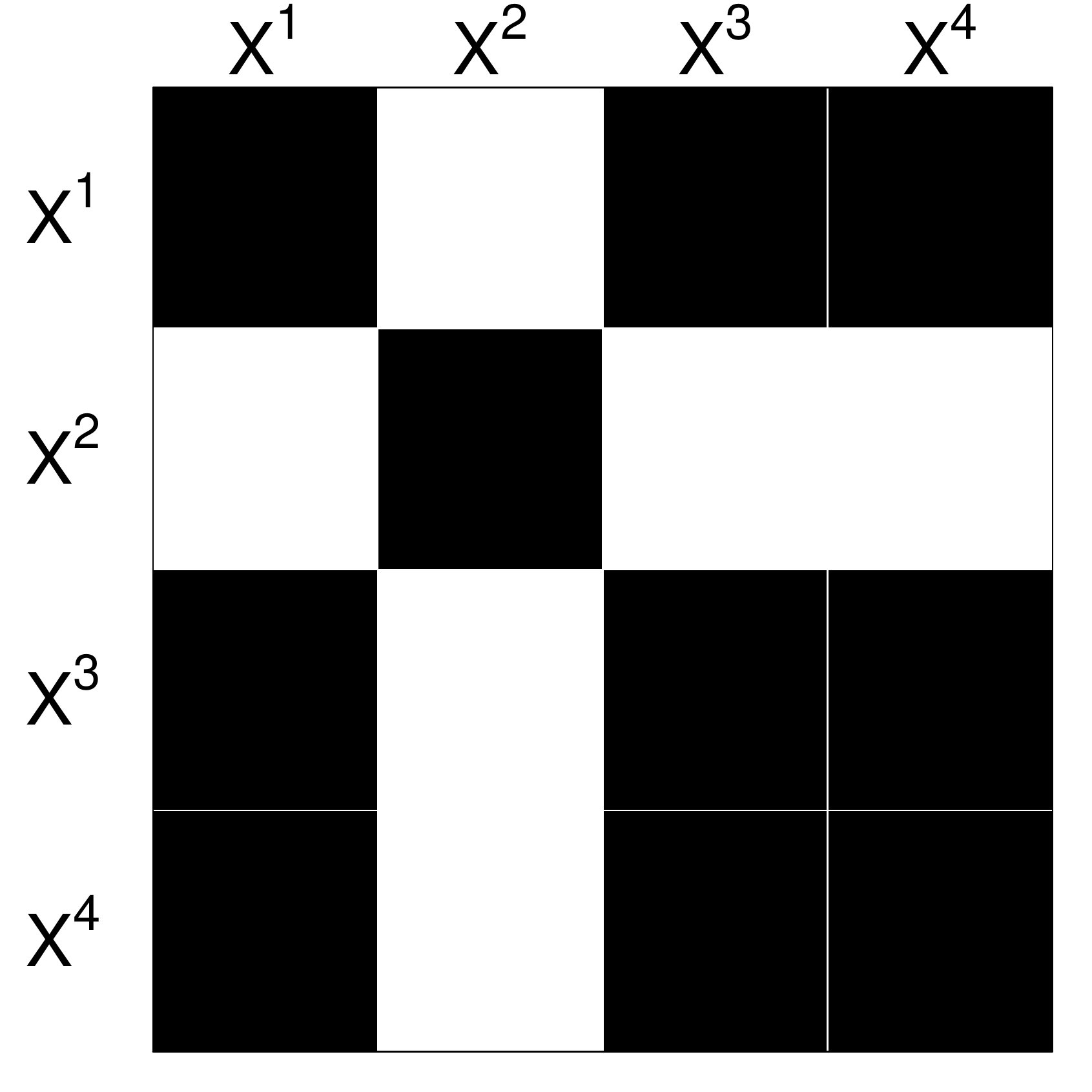}\\
\centering (b) \\
\centering \includegraphics[scale=0.16]{base_sigma.pdf} 
\centering \includegraphics[scale=0.16]{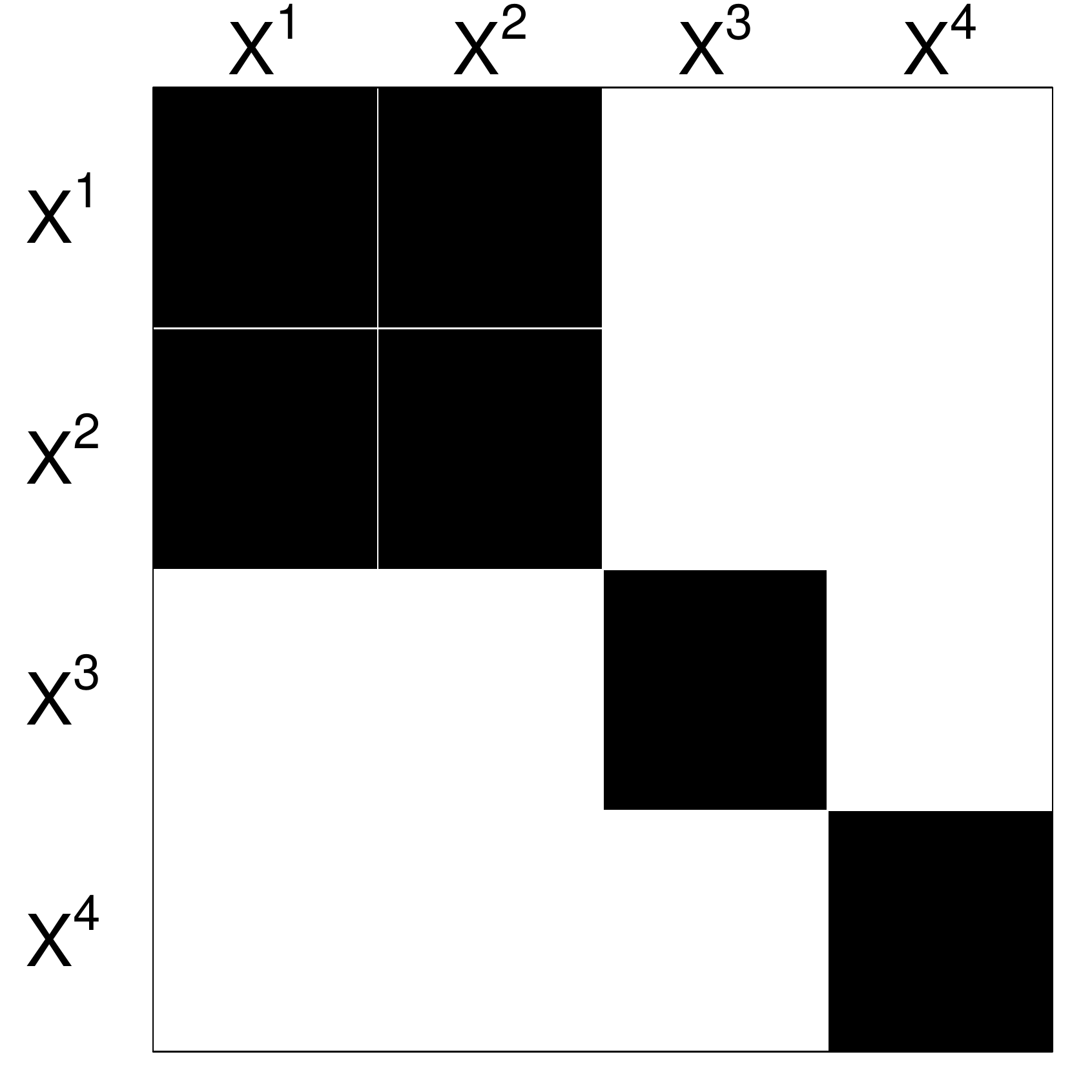}  
\centering \includegraphics[scale=0.16]{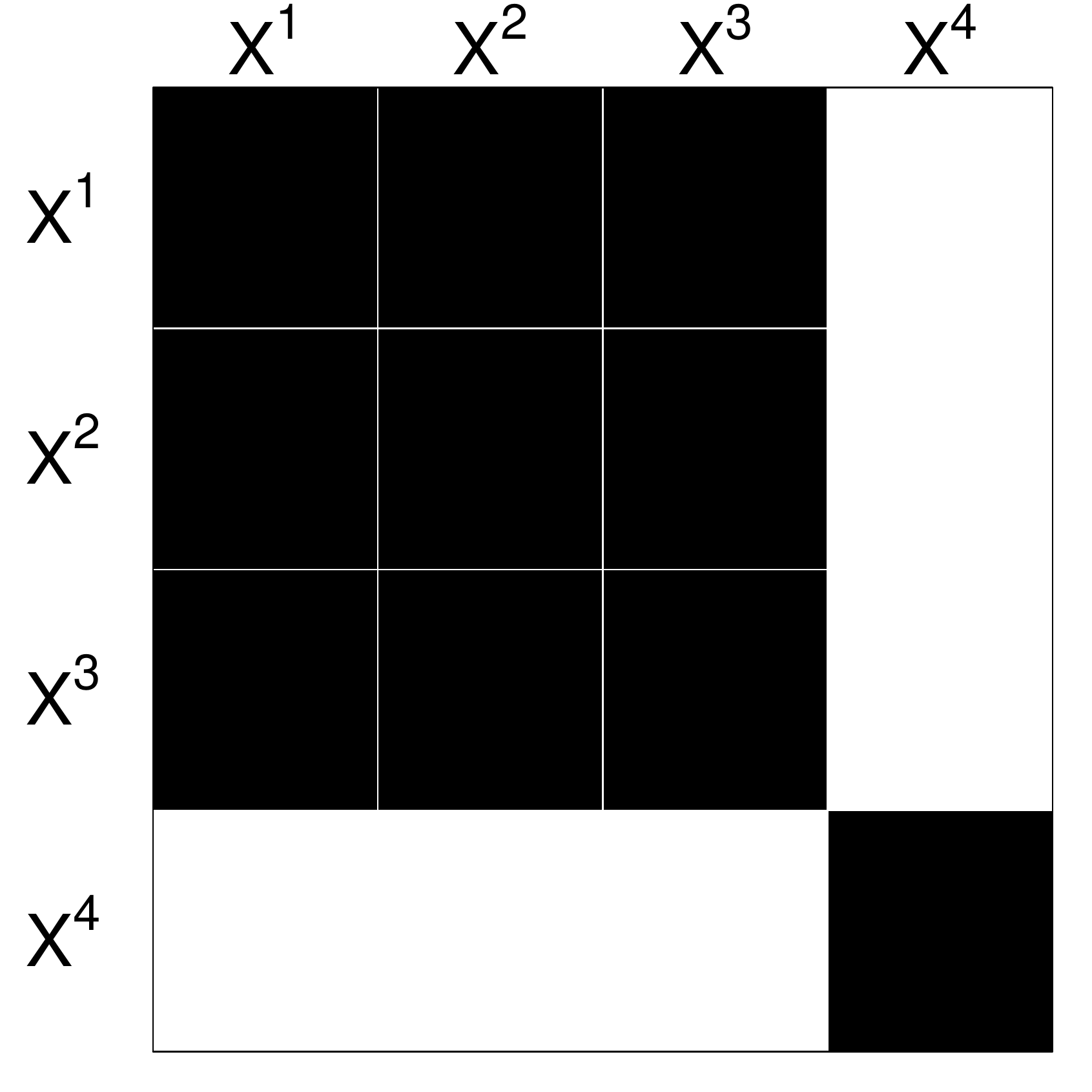} 
\centering \includegraphics[scale=0.16]{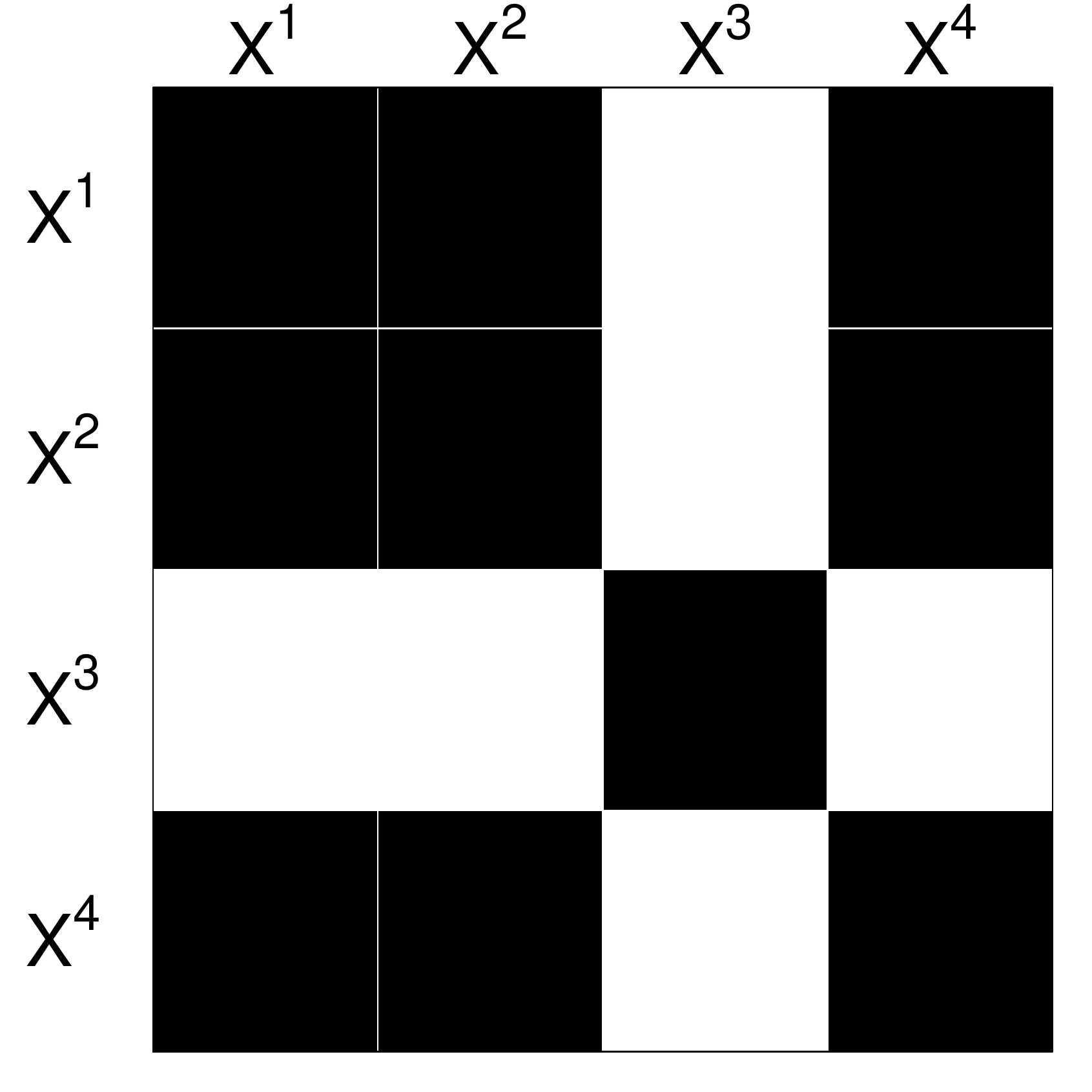}\\
\centering (c) \\
\end{minipage} 
\caption{Example of the support of $ \Sigma^{[q]}$ in a case of four variables.  If the variable of  row $j$ and the variable of column $j'$ are in the same block then cell $(j,j')$ is painted in black. This cell is painted in white otherwise. (a) Graphical representation of $\boldsymbol{\sigma}_k^{[q]}=(\{1,2\},\{3,4\})$; (b) Elements of $ \Sigma^{[q]}$ if $b_{from}^{[q]}=1$; (c) Elements of  $ \Sigma^{[q]}$ if $b_{from}^{[q]}=2$.\label{voisinsigma}}
\end{figure}

A possible deterministic neighborhood of $\boldsymbol{\sigma}^{[q]}$ could be defined as the set of models where, at most one variable is affected, for one component, in another block (possibly creating a new block): $\left\{ \boldsymbol{\sigma} : \exists ! (k,b,j) \; j \in \boldsymbol{\sigma}_{kb}^{[q]} \text{ and } j \notin \boldsymbol{\sigma}_{kb} \right\} \cup \left\{\boldsymbol{\sigma}^{[q]} \right\}$. However, as this deterministic neighborhood can be very large, our proposal distribution allows reducing it to a stochastic neighborhood  $\Sigma^{[q]}$ by reducing the number of $(k,b)$ where $\boldsymbol{\sigma}_{kb}$ could be different to $\boldsymbol{\sigma}_{kb}^{[q]}$. Thus, one component $k^{[q]}$ is randomly sampled in $\{1,\ldots,g\}$ then one block $b_{from}^{[q]}$ is randomly sampled in $\{1,\ldots,B_{k^{[q]}}^{[q]}\}$. Another block $b^{[q]}$  is randomly sampled in $\{1,\ldots,B_{k^{[q]}}^{[q]}\} \setminus b_{from}^{[q]}$ and the set $b_{to}^{[q]}=\{b^{[q]},B_{k^{[q]}}^{[q]}+1\}$ is built. The stochastic neighborhood $\Sigma^{[q]}$ is then defined as:
\begin{small}
\begin{equation}
\Sigma^{[q]} = \left\{ \boldsymbol{\sigma} : \exists ! (k,b,j) \; j \in \boldsymbol{\sigma}_{kb}^{[q]}, \; j \notin \boldsymbol{\sigma}_{kb} \text{ and } j \in \boldsymbol{\sigma}_{kb'}  \text{ with } k=k^{[q]}, \; b=b_{\text{from}}^{[q]}, \; b' \in b_{\text{to}}^{[q]} \right\} \cup \left\{\boldsymbol{\sigma}^{[q]} \right\}.
\end{equation}
\end{small}
 We denote the elements of $\Sigma^{[q]} $ as $ \boldsymbol{\sigma}^{[q+\varepsilon(e)]} $ where $\varepsilon(e)=\frac{e}{| \Sigma^{[q]} |+1}$ and $e=1,\ldots,| \Sigma^{[q]} |$. Figure~\ref{voisinsigma} shows an illustration of this definition.

At the generation pattern step, the previous algorithm needs the value of $p(\textbf{x}|g,\boldsymbol{\sigma})$ $\forall \boldsymbol{\sigma} \in  \Sigma^{[q]}$. By using the BIC approximation \citep{Sch78,Leb06}, this probability is approximated by:
\begin{equation}
\ln p(\textbf{x}|g,\boldsymbol{\sigma}) \simeq L(\hat{\boldsymbol{\theta}};\textbf{x},g,\boldsymbol{\sigma})-\frac{\nu_{\text{\textsc{ccm}}}}{2}\log(n),
\end{equation}
$\hat{\boldsymbol{\theta}}$ being the maximum likelihood estimator obtained by the GEM algorithm previously described in Section~\ref{estimation}. Thus, at the iteration $[q]$, for each $e=1,\ldots,| \Sigma^{[q]} |$,  estimator $\hat{\boldsymbol{\theta}}^{[q+\varepsilon(e)]}$ associated to element $\boldsymbol{\sigma}^{[q+\varepsilon(e)]}$ is computed by the GEM algorithm.

\paragraph*{Initialization:} Whatever the initial value selected for $\boldsymbol{\sigma}^{[0]}$, the algorithm converges to the same value of $\boldsymbol{\sigma}$. However, this convergence can be very slow when the initialization is poor. Since blocks consist of the most correlated variables, a Hierarchical Ascendant Classification (HAC) is used on the matrix of Cramer's V distances on the variable couples. The partition produced by the HAC minimizing the block number without blocks consisting of more than four variables is chosen for each $\boldsymbol{\sigma}^{[0]}_k$. For the initialization, the variables number of a block is limited to four, because very few blocks having more than four variables were exhibited in the course of our experiments. Obviously, the Gibbs algorithm can then violate this initial constraint if necessary.

\paragraph*{Stopping criterion:} The algorithm is stopped when $q_{\max}$ successive iterations have not discovered a better model.

\subsection{Consequences of the Gibbs algorithm on the GEM algorithm \label{MCMCapprox}}	
\paragraph*{Initialization of the GEM algorithm: } 
At iteration $[q]$ of the Gibbs algorithm, the GEM algorithm estimates $\hat{\boldsymbol{\theta}}^{[q+\varepsilon(e)]}$ associated to model $\boldsymbol{\sigma}^{[q+\varepsilon(e)]}$ for  $e=1,\ldots,| \Sigma^{[q]} |$. Since these models are close to $\boldsymbol{\sigma}^{[q]}$, their maximum likelihood estimators should be closed to $\hat{\boldsymbol{\theta}}^{[q]}$. The GEM algorithm initialization is also done by the value of $\hat{\boldsymbol{\theta}}^{[q]}$ for the not modified blocks. Thus, $\boldsymbol{\theta}_{kb}^{[q+\varepsilon(e)](0)}=\hat{\boldsymbol{\theta}}_{kb}^{[q]}$ if the blocks are not modified ($\boldsymbol{\sigma}_{kb}^{[q+\varepsilon(e)]}=\boldsymbol{\sigma}_{kb}^{[q]}$). For the other blocks, the continuous parameters are randomly sampled. For those blocks, in order to avoid the combinatorial problems, we use a sequential method to initialize $\boldsymbol{\delta}_{kb}^{[q+\varepsilon(e)](0)}$: the surjections from $\boldsymbol{x}^{\{kb\}1} $ to $\boldsymbol{x}^{\{kb\}j} $ are sampled, according to $\textbf{x}$ and to the continuous parameters previously sampled $(\rho_{kb}^{[q+\varepsilon(e)](0)},\boldsymbol{\alpha}_{kb}^{[q+\varepsilon(e)](0)},\boldsymbol{\tau}_{kb}^{[q+\varepsilon(e)](0)}) $, for each $j=2,\ldots,d^{\{kb\}}$ as follows:
\begin{equation}
\begin{small}
\boldsymbol{\delta}_{kb}^{.j[q+\varepsilon(e)](0)} \propto \prod_{i=1}^{n} p(x_i^{\{kb\}1},x_i^{\{kb\}j};\rho_{kb}^{[q+\varepsilon(e)](0)},\boldsymbol{\alpha}_{kb}^{1[q+\varepsilon(e)](0)},\boldsymbol{\alpha}_{kb}^{j[q+\varepsilon(e)](0)},\boldsymbol{\tau}_{kb}^{[q+\varepsilon(e)](0)},\boldsymbol{\delta}_{kb}^{.j})^{z_{ik}^{[q]}}, \label{initdelta}
\end{small}
\end{equation}
$ \text{ where }  \boldsymbol{\delta}_{kb}^{.j[q+\varepsilon(e)]}=(\boldsymbol{\delta}_{kb}^{hj[q+\varepsilon(e)]};h=1,\ldots,m_1^{\{kb\}})$ and where $z_{ik}^{[q]}=E \big[Z_{ik}  | \boldsymbol{x}_i, \boldsymbol{\theta}^{[q]} \big]$.
	
\paragraph*{Remark about $r_{\max}$: }
As said in Section \ref{EMglo}, the algorithm is stopped after a fixed number of iterations $r_{\max}$. If the algorithm is stopped before its convergence, the proposed initialization limits the problems. Indeed, if the model has a high \emph{a posteriori} probability, it will stay in  neighborhood $\Sigma^{[q]}$ during some successive iterations, so its log-likelihood will increase.

\section{Simulations \label{simulations}}
Table~\ref{valparam} presents the adjustment parameters values used for all the simulations. 

\begin{table}[h!]
\begin{center}
\begin{tabular}{ccccc}
\hline Algorithms & Gibbs & GEM & Metropolis-Hastings & EM \\ 
\hline Criteria & $q_{\max}=20\times d$ &$r_{\max}=10$ & $s_{\max}=1$ & $t_{\max}=5$ \\ 
\hline 
\end{tabular} 
\caption{Values of the different stopping criteria.\label{valparam}}
\end{center}
\end{table}

As these algorithms are interlocked, the iterations number of the most internal algorithms are small. Since the number of possible models increases with $d$, we propose to set: $q_{\max}=20\times d$. When the best model is selected by the Gibbs algorithm, this latter will stay in this model during many iterations so the Metropolis-Hastings and the EM algorithm are performed many times. Thus, it is not necessary to have a large number of iterations as stopping criterion.

\subsection{Study of the algorithm for the $\boldsymbol{\delta}_{kb}$ estimation}

In this section, we illustrate the performance of the Metropolis-Hastings algorithm used for the $\boldsymbol{\delta}_{kb}$ estimation (see Section~\ref{discestim}) and the relevance of its initialization (see Equation~\eqref{initdelta}). Since this algorithm is interlocked in the Gibbs and in the GEM algorithm, we need it to converge quickly. The following simulations show that the algorithm stays relevant up to six modalities per variable and up to six variables per block. These conditions hold in most  situations.

Samples of size 200 described by variables having the same number of modalities are generated by a  mixture between an independence distribution and a maximum dependency distribution. The parameter estimation is also performed by the Metropolis-Hastings algorithm, described in Section~\ref{discestim}, since only one class is generated. The discrete parameters initializations are performed according to Equation~\eqref{initdelta} with $z_{i1}=1$ for $i=1,\ldots,200$.

\begin{figure}[h!]
\begin{minipage}{0.49\textwidth}
\begin{center}
\centering \includegraphics[scale=0.35]{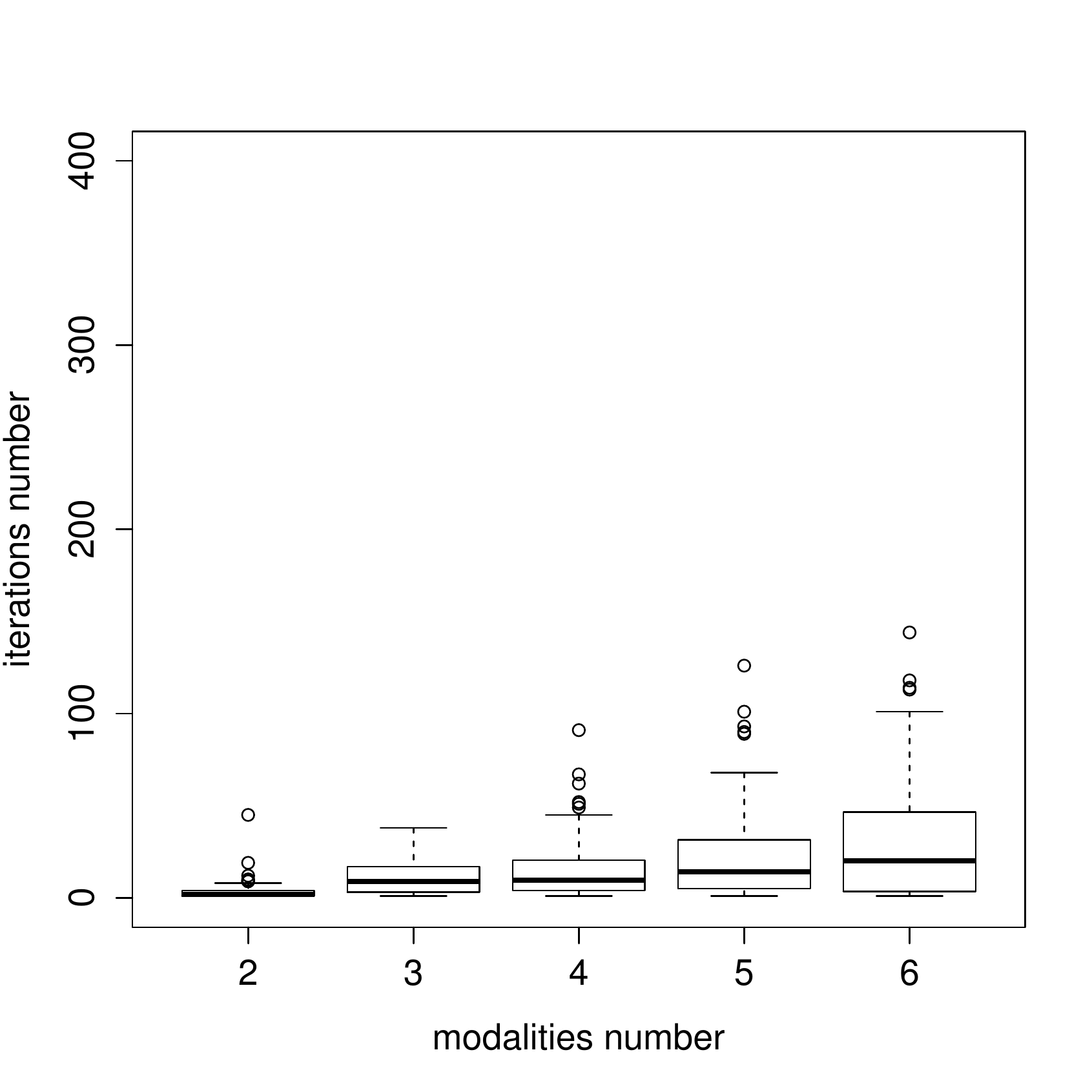} \\(a)\\
\centering \includegraphics[scale=0.35]{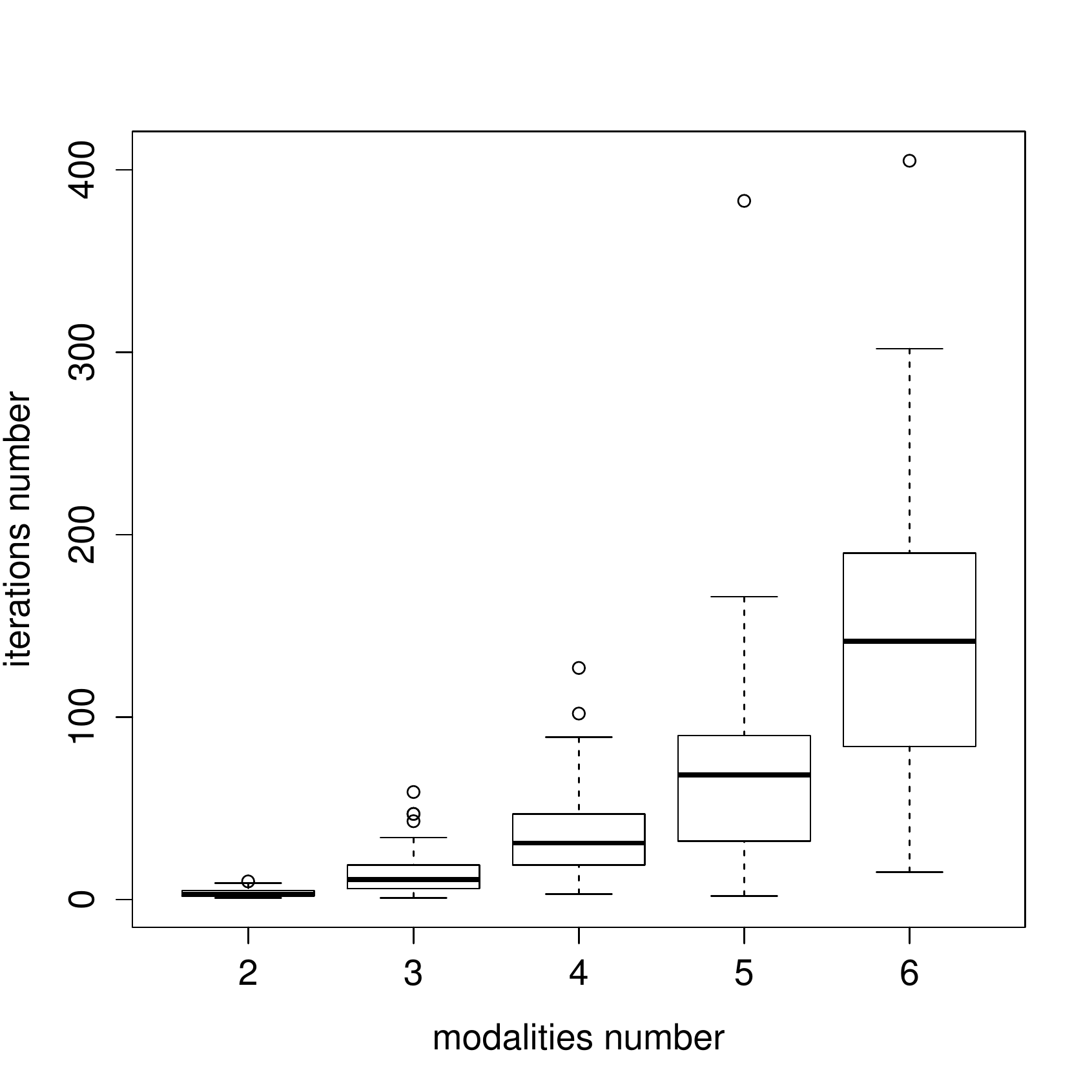} \\(c)\\
\end{center}
\end{minipage}
\begin{minipage}{0.49\textwidth}
\begin{center}
\centering \includegraphics[scale=0.35]{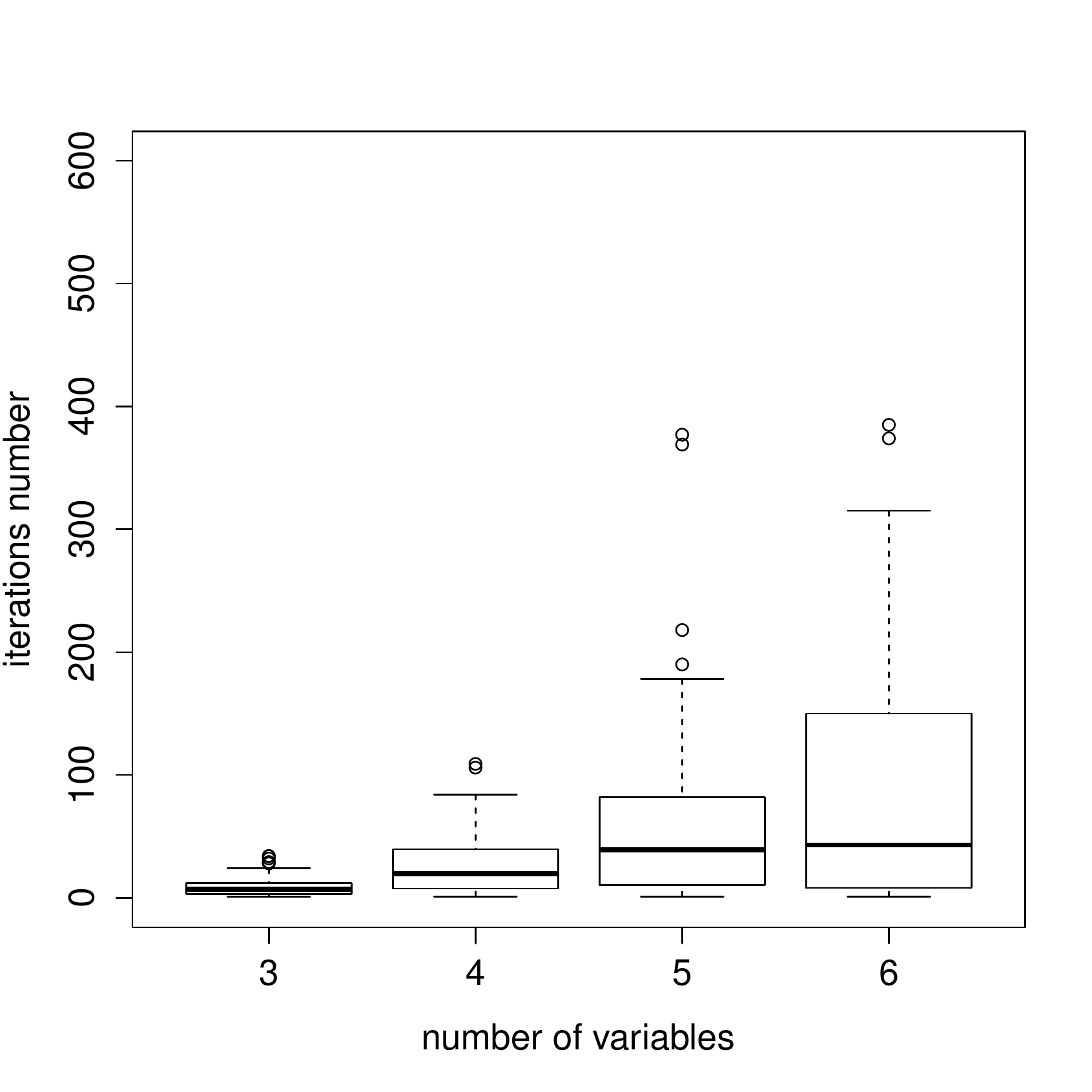} \\(b)\\
\centering \includegraphics[scale=0.35]{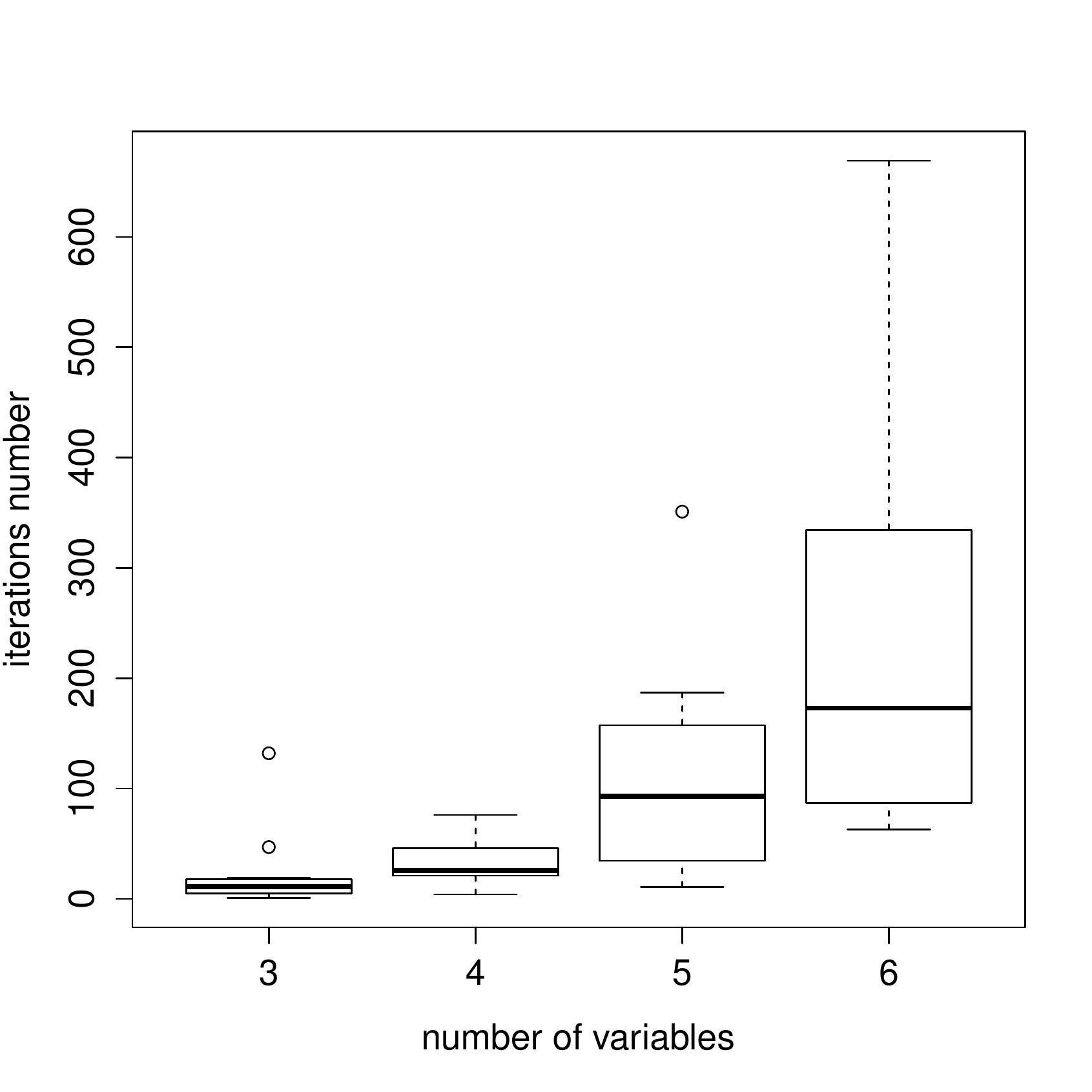} \\(d)\\
\end{center}
\end{minipage}
\caption{{\footnotesize Box-plots of the number of  iterations required by the Metropolis-Hastings algorithm for finding the best links between modalities, according to the number of modalities when datasets are simulated with a proportion of maximum dependency distribution equal to 0.5. (a) Three variables with the proposed initialization; (b) Three modalities per variable  with the proposed initialization; (c) Three variables with a random initialization; (d)  Three modalities per variable  with a random initialization.\label{simuldelta}}}
\end{figure}

Figure~\ref{simuldelta} shows the box-plots of the iterations number needed by the Metropolis-Hastings algorithm for finding the true links between modalities maximizing the likelihood\footnote{In fact, the algorithm is stopped as soon as it finds a discrete estimator involving a likelihood higher than or equal to the likelihood obtained with the true discrete parameters used for the simulation.}. 
According to these simulations, one observes that the results of this algorithm are good as a result of its initialization which allows significantly reducing the number of iterations needed in order to find the maximum likelihood estimators.
 
\subsection{Study of the algorithm for model selection}
In order to illustrate the efficiency of the algorithm for the model selection (and also the included estimation process),
we want to study the evolution of the Kullback-Leibler divergence according to the number of variables and to the size of the data set. Thus, 100 samples are generated for many situations according to the \textsc{ccm} with two components. Note that the parameter $u$ is introduced for controlling the overlapping of classes: when it is close to one their overlapping (Bayes error) is close to one. These parameters set the error rate to 0.10 for each studied situation:
\begin{small}
$$\boldsymbol{\sigma}_{kb}=(d/b,1+d/b) \quad \rho_{kb}=0.6(1-u) \quad \boldsymbol{\tau}_{kb}=(0.60,0.20,0.20), $$
$$\delta_{1b}^{h2h'}=1 \text{ iff } h=h' \quad \delta_{1b}^{122}=\delta_{1b}^{223}=\delta_{1b}^{321}=1 \quad \boldsymbol{\alpha}_{1b}^j=(0.20,0.20,0.60),  $$
$$  \boldsymbol{\alpha}_{2b}^1=\boldsymbol{\alpha}_{1b}^1(1-u) + (0.075,0.850,0.075)u \quad \text{and} \quad  \boldsymbol{\alpha}_{2b}^2=\boldsymbol{\alpha}_{1b}^2(1-u) + (0.850,0.075,0.075)u.$$
\end{small}
Table~\ref{simulkull} shows the mean and the standard deviation of the Kullback-Leibler divergence between the parameters used for the dataset generation and the estimated parameters according to the number of variables. When $n$ increases, the Kullback-Leibler divergence converges to zero. It confirms the good behavior of the proposed algorithm.
\begin{table}[h!]
\begin{center}
\begin{tabular}{ccccc}
\hline $d \setminus n$ & 100 & 200 & 400 &800 \\ 
\hline  4 & \textbf{0.77} (\emph{1.34})&\textbf{0.26} (\emph{0.26}) & \textbf{0.15} (\emph{0.05})&\textbf{0.12} (\emph{0.05}) \\
\hline  6 &\textbf{1.22} (\emph{1.77}) &\textbf{0.27} (\emph{0.14}) & \textbf{0.09} (\emph{0.07})& \textbf{0.05} (\emph{0.05})\\
\hline  8 &\textbf{1.72} (\emph{2.50}) &\textbf{0.41} (\emph{0.20}) & \textbf{0.09} (\emph{0.05})& \textbf{0.05} (\emph{0.03})\\
\hline  10 &\textbf{1.73} (\emph{4.06}) & \textbf{0.52} (\emph{0.14})& \textbf{0.10} (\emph{0.03})&\textbf{0.04} (\emph{0.03}) \\
\hline
\end{tabular}
\caption{\textbf{Mean} (\emph{standard deviation}) of the Kullback-Leibler divergence.\label{simulkull}}
\end{center}
\end{table}

\section{Application \label{applications}}

\subsection{Dentistry clustering}

	The Handelman's dentistry data \citep{Han86} display the evaluation of 3869 dental x-rays (sound or carious) performed by five dentist and possibly showing incipient caries. This data set has been clustered by several models in the past. It is suggested that there are two main classes: the sound teeth and the carious ones.

	According to the BIC criterion, data are split into three classes by \textsc{cim}. Furthermore, dependencies are observed between the variables into classes since the Cramer's V computed per class is not close to zero. Thus, \citet{Esp89} apply a  log-linear mixture model to fit the data. The authors set the model, by adding some assumptions to better fit the data. More precisely, they consider a mixture with four components. The first two components take into account the interactions between dentists 3 and 4. The last two components are specific since they allow only one modality interaction, when all the diagnoses are respectively carious and sound. Note that these assumptions are required by the above authors due to their realistic nature. Indeed, this model fits the data better than \textsc{cim}. On the other hand, its interpretation needs the analysis of four classes.

	As the last two classes seem artificial, \citet{Qu96} prefer to use the random effects models in a latent class analysis with two classes. They assume  that the conditional dependencies can be modeled by a single continuous latent variable varying among the individuals. According to the authors, one class represents the sound teeth and the other represents the carious ones, while the random effect represents all the patient-specific unrecorded characteristics of the x-ray images. Their model does not need the two additional artificial classes. Thus their interpretation is easier.

	We now display the results of the proposed model \textsc{ccm} estimated with the R package \texttt{Clustericat} (the code is presented in Appendix~\ref{tuto}). The BIC criterion selects two classes with a value of -7473. It claims that \textsc{cmm} better fits the data than the model of \citet{Qu96} since their BIC criterion value is -7487. The BIC criterion values for \textsc{cim} and \textsc{cmm} are displayed in Table~\ref{bicdent}. We indicate the computing time (in seconds), obtained with an Intel Core i5-3320M processor, to estimate \textsc{cmm} where 20 MCMC chains were started with a stopping rule $q_{\max}=100$ while \textsc{cim} needs less than 0.1 sec with the R package RMixmod \citep{Leb12}.

\begin{table}[h!]
\begin{center}
\begin{tabular}{cccccc}
\hline \multicolumn{2}{c}{$g$} & 1 & 2 & 3 & 4  \\ 
\hline  \textsc{cim}&BIC&-8766 & -7511 & \textbf{-7481} & -7503 \\ 
 \textsc{ccm}&BIC& -7743 & \textbf{-7473} & -7481 & -7503\\ 
&time (sec) & 1.7 & 4.9 & 6.1 & 7.7\\ 
\hline 
\end{tabular}
\caption{BIC criterion values for the \textsc{cim} and the \textsc{cmm} according to different numbers of classes for the dentistry data set. For each model, the best results according to the BIC criterion are in bold. The computing time in seconds is indicated for \textsc{cmm} where 20 MCMC chains were started with a stopping rule $q_{\max}=500$.\label{bicdent}} 
\end{center}
\end{table}

We note that \textsc{cmm} obtains better values for the BIC criterion than \textsc{cim} when $g=1,2$. When the number of classes is larger ($g\geq 3$) the best model of \textsc{cmm} assumes the conditional independence between variables.

The BIC criterion selects two classes for \textsc{cmm} and this is coherent with a clustering of the teeth between the sound and the carious ones. Furthermore, the two main characteristics of the model fixed by \citet{Esp89} are automatically detected by the model: importance of the two modality crossings where all the dentists have the same diagnosis and a dependency between the diagnosis of  dentists 3 and 4. Thus, the estimated model is coherent with the imposed model of \citet{Esp89} while no information was given \emph{a priori}.

The fitted model can be interpreted as:

	\begin{itemize}
	\item the majority class ($\pi_1=0.86$)  mainly gathers the sound teeth. There is a strong dependency between the five dentists ($\boldsymbol{\sigma}_1=(\{1,2,3,4,5\})$ and $\rho_{11}=0.35$). The dependency structure of the maximum dependency distribution indicates an over-contribution of both modality interactions where the five dentists have the same diagnosis, especially when they claim that the teeth are sound ($\tau_{11}^{\text{all\_sound}}=0.93$ and $\tau_{11}^{\text{all\_carious}}=0.07$). This class could be interpreted as grouping the teeth who these diagnosis is obvious.
	\item the minority class ($\pi_2=0.14$) groups principally the carious teeth. There is a dependency between dentists 3 and 4 while the diagnosis of the other ones are independent given the class ($\boldsymbol{\sigma}_2=(\{3,4\},\{1,2,5\})$, $\rho_{21}=0.31$ and $\rho_{22}=0$). This class could be interpreted as grouping the teeth whose diagnosis is more complex.
	\end{itemize}

\begin{figure}[h!]
\centering \includegraphics[scale=0.6]{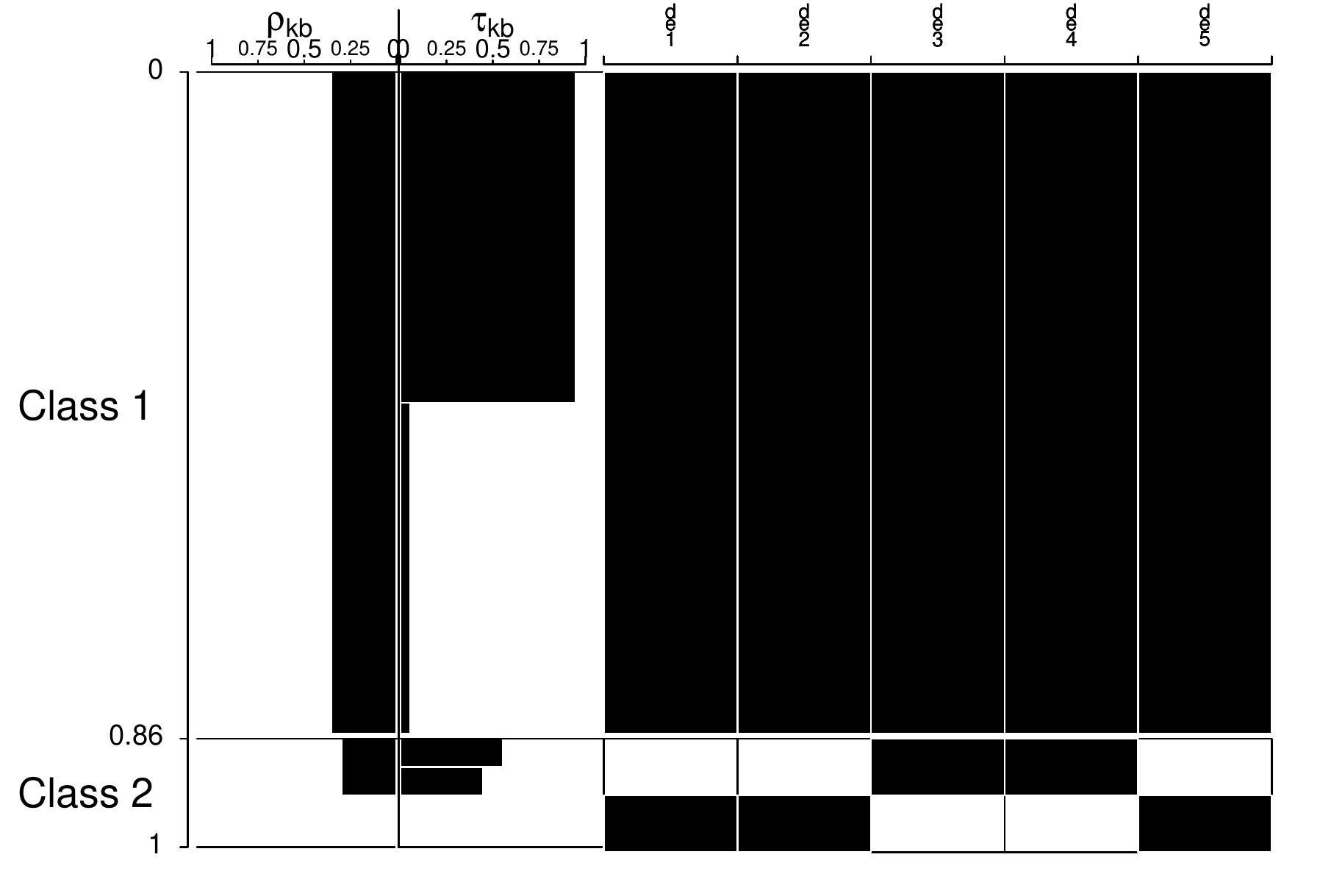} 
\caption{Summary of the best \textsc{ccm} according to BIC for the dentists data set. \label{resumdentists}} 
\end{figure}

Figure \ref{resumdentists} helps the interpretation of the clusters for the \textsc{ccm} with two components (best model according to the BIC criterion). On ordinates, the estimated classes are represented with respect to their proportions in decreasing order. Their corresponding area depends on their proportion. The cumulated proportions are indicated on the left side. On abscissae, three indications are given. The first one is the inter-variables correlations ($\rho_{kb}$) for all the blocks of the class ordered by their strength of correlation (in decreasing order). The second one is the intra-variables correlations ($\boldsymbol{\tau}_{kb}$) for each block drawn according to their strength dependencies (in decreasing order). The third is the variables repartition per block. A black cell indicates that the variable is assigned to the block and a white cell indicates that, conditionally on this class, the variable is independent of the variables of this block. For example, this figure shows that the first class has a proportion of $0.86$ and that all the variables are assigned into the same block.

\subsection{Calves clustering}
In this section, the results obtained by the \textsc{ccm} are compared to those obtained for the \textsc{cim} by the RMixmod software \citep{Leb12}.  The ``Genes Diffusion'' company has collected information from French breeders in order to cluster calves. The 4270 studied calves are described by nine variables related to behavior (aptitude for sucking \emph{Apt}, behavior of the mother just before the calving \emph{Iso}) and health (treatment against omphalitis \emph{TOC}, respiratory disease \emph{TRC} and diarrhea \emph{TDC}, umbilicus disinfection \emph{Dis}, umbilicus emptying \emph{Emp}, mother preventive treatment against respiratory disease \emph{TRM} and diarrhea \emph{TDM}).

Table~\ref{veauind} displays the BIC criterion values and the number of parameters for the \textsc{cim} and \textsc{ccm} models. Furthermore, the computing time in minutes (obtained with an Intel Core i5-3320M processor) to estimate \textsc{ccm} by starting 20 MCMC chains with a stopping criterion of $q_{\max}=180$ while \textsc{cim} needs 3 seconds with the R package RMixmod \citep{Leb12}.

\begin{figure}[h!]
\centering \includegraphics[scale=0.6]{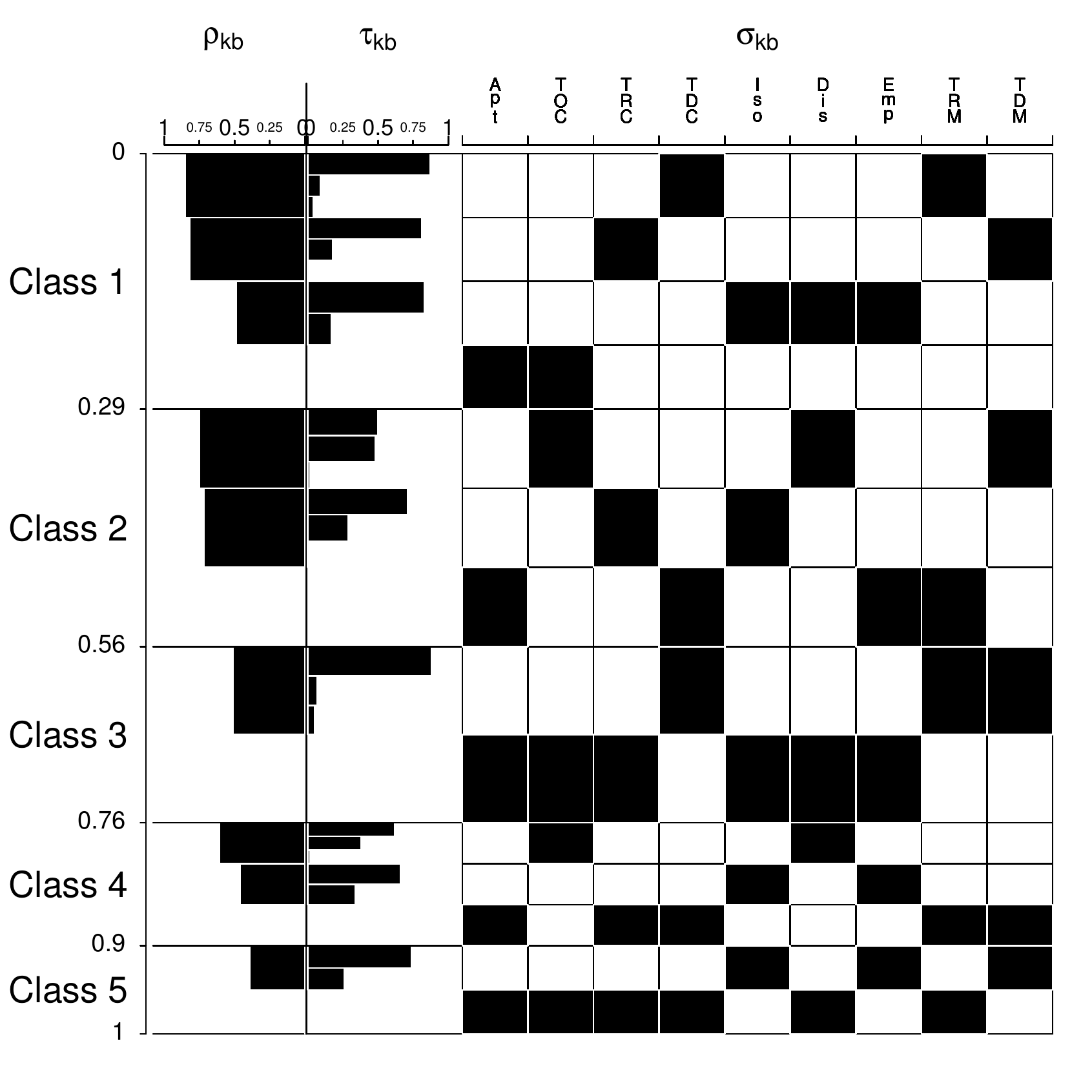} 
\caption{Summary of the best \textsc{ccm} according to BIC for the calves data set. \label{resumGD}} 
\end{figure}

\begin{table}[h!]
\begin{center}
\begin{footnotesize}
\begin{tabular}{cccccccccc}
\hline \multicolumn{2}{c}{$g$} & 1 & 2 & 3 & 4 & 5 & 6 & 7 & 8 \\ 
\hline  \textsc{cim}&BIC& -28589 & -26859 & -26526 & -26333 & -26238 & -26235 & -26226 & \textbf{-26185 } \\ 
&$\nu_{\textsc{cim}}$ & 17 & 35 & 53 & 71 & 89 & 107 & 125 &\textbf{ 143 } \\ 
\hline \textsc{ccm}&BIC& -26653 & -26289 & -26173 & -26038 & \textbf{-26025} & -26059 & -26045 &-26058\\ 
&$\nu_{\textsc{ccm}}$& 24 & 48 & 80& 89 & \textbf{112}&131 &148&163 \\ 
&time (min)&0.97 & 3.32 & 6.16 & 6.56& \textbf{10.03}& 11.76 & 12.31 & 14.92\\ 
\hline 
\end{tabular}
\caption{Results for the \textsc{cim} and the \textsc{cmm} according to different class numbers. For both models, first row corresponds to the BIC criterion values and the second row indicates the continuous parameter number. For each model, the best results according to the BIC criterion are in bold. The computing time for the \textsc{ccm} estimation is given in minutes.\label{veauind}} 
\end{footnotesize}
\end{center}
\end{table}
For the \textsc{cim}, the BIC criterion selects a high number of classes, since it selected eight classes. The interpretation of the clusters is also difficult and we can assume that the estimator's quality is very poor. Figure \ref{resumGD} helps the interpretation for the \textsc{ccm} with five components (best model according to the BIC criterion). Its interpretation is the same as the interpretation of Figure~\ref{resumdentists}. For example, this figure shows that the first class has a proportion of $0.29$ and that it is composed of four blocks. The most correlated block of the first class has $\rho_{kb}\simeq 0.80$ and the strength of the biggest modalities link is also close to $0.85$. This block consists of the variables \emph{TDC} and \emph{TRM}.
Here is now a possible interpretation of Class~1 (note that the others classes are also meaningful; see details in \citet{Mar13}):
\begin{itemize}
\item \textbf{General: } This class has a proportion equal to 0.29 and consists of three blocks of dependency and one block of independence.
\item \textbf{Block 1: } There is a strong correlation ($\rho_{11}$) between the variables diarrhea treatment of the calf and mother preventive treatment against respiratory disease, especially between the modality no treatment against the calf diarrhea and the absence of preventive treatment against respiratory disease of its mother ($\boldsymbol{\tau}_{11}$ and $\boldsymbol{\delta}_{11}$).
\item \textbf{Block 2: } There is a strong correlation ($\rho_{12}$) between the variables treatment against respiratory illness of the calf and mother preventive treatment against diarrhea, especially between the modality preventive treatment against respiratory illness of the calf and the presence of diarrhea preventive treatment  of its mother ($\boldsymbol{\tau}_{12}$  and $\boldsymbol{\delta}_{12}$).
\item \textbf{Block 3: } There exists another strong link  between the behavior of the mother, the emptying of the umbilical and its disinfection ($\boldsymbol{\tau}_{13}$  and $\boldsymbol{\delta}_{13}$).
\item \textbf{Block 4: } This block is characterized by absence of preventive treatment against omphalitis and having $50\%$ of the calves infected by this illness ($\boldsymbol{\alpha}_{14}$).
\end{itemize}

\section{Conclusion \label{conclu}}
By using the block extension of the latent class model, a new mixture model is proposed for clustering categorical data by taking into account the intra-class correlation. The block distribution is defined as a mixture between an independent distribution and a maximum dependency distribution. This specific distribution, which remains parsimonious, is compared to the full latent class model and allows different levels of interpretation. The blocks of variables detect the conditional dependencies between variables while their strengths are reflected by the proportions of maximum dependency distribution. The parameters of the block distribution reflect the links and the strength between modalities.

The parameter estimation and the model selection are simultaneously performed via a Gibbs sample-type algorithm. It allows reducing the combinatorial problems of the block structure detection and the links between modalities search for the estimation of the maximum dependency distribution. The results are good when the number of modalities is small for each variable. For more than six modalities, the detection of other links encounters some persistent difficulties. So the algorithm can be slow in this case. The proposed approach to estimate the block structure is not adapted for data sets with many variables. A deterministic but sub-optimal solution could be used to perform a forward algorithm.

The R package \emph{Clustericat} allows clustering categorical data sets by using \textsc{cmm}. This package is available on Rforge at the following url \emph{https://r-forge.r-project.org/R/?group\_id=1803}. 

The proposed model can be easily extended to the case of ordinal data. For this purpose, some additional constraints on the dependency structure of each distribution of maximum dependency need to be added.

\paragraph*{Acknowledgments: } The author thank the Associate Editor and the three anonymous referee for their useful comments and references. The authors are grateful to the Genes Diffusion corporation for the provision of the data set and especially its members: Am\'elie Vall\'ee, Julie Hamon and Claude Grenier. We are grateful to Parmeet Bhatia, Modal Team engineer, and  St\'ephane Chr\'etien for their precious assistance. This work was financed by DGA and Inria.

\bibliography{biblio}
\bibliographystyle{authordate1}

\begin{appendix}
\section{Dentistry clustering with the R package \texttt{Clustericat}} \label{tuto}
The R package \texttt{Clustericat} is available on Rforge website at the following url: \emph{https://r-forge.r-project.org/R/?group\_id=1803}. This section presents the code used to cluster the dentistry data set.\bigskip\\
\# Loading of the data set\\
$>$ \texttt{data("dentist")} \bigskip\\
\# to define the parameters of the algorithm performing the estimation\\
\# here 25 MCMC are performed with a stopping criterion equals to \\
\# 200 successive iterations having not found a better model\\
$>$ \texttt{st $<$- strategycat(dentist, nb\_init=25, stop\_criterion=200)}\bigskip\\
\# estimation of the model for a class number equal to 2.\\
\# for the data set with five binary variables (modal)\\
$>$ \texttt{res $<$- clustercat(dentist, 2, modal=rep(2,5), st)}\bigskip\\
\# presentation of the best model\\
$>$ \texttt{summary(res)}\bigskip\\
\# presentation of the parameters of the conditional dependencies for the best model\\
$>$ \texttt{summary\_dependencies(res)}\bigskip\\
\# a plot summarizing the best model like Figure~\ref{resumdentists}\\
$>$ \texttt{plot(res)}
\end{appendix}

\end{document}